\documentclass[10pt,oneside,twocolumn,aps,pra,preprintnumbers,bibnotes,superscriptaddress,amsmath,amssymb]{revtex4-2}

\usepackage{caption}
\usepackage[utf8]{inputenc}
\usepackage{graphicx} 
\usepackage{subcaption}
\usepackage{float}
\usepackage{epstopdf}
\usepackage{amsmath}
\usepackage{mathbbol}
\usepackage{amssymb}
\usepackage{amsfonts}

\usepackage{amsthm}
\usepackage{mathtools}
\usepackage{bm}
\usepackage{placeins}
\usepackage{commath}
\usepackage{braket}
\usepackage{physics}
\usepackage{cancel}
\usepackage[usenames,dvipsnames]{xcolor}
\usepackage{ulem}
\usepackage{hyperref}
\usepackage{lipsum}
\usepackage{comment}
\usepackage{tikz}
\usepackage{tikzscale}
\usetikzlibrary{quantikz2,calc,arrows,chains,matrix,positioning,scopes}

\usepackage{letltxmacro}
\newcommand{\id}{\mathbb{I}}

\begin{document}

\title{Sequential strategy for depolarizing strength estimation}

\author{Yasir Mehmood}
\affiliation{Università degli Studi di Napoli Federico II, Via Cinthia 21, 80126 Napoli, Italy.}
\affiliation{Istituto Nazionale di Ottica del Consiglio Nazionale delle Ricerche (CNR-INO), Largo Enrico Fermi 6, I-50125 Firenze, Italy.}

\author{Enrico Trombetti}
\affiliation{Università degli Studi di Napoli Federico II, Via Cinthia 21, 80126 Napoli, Italy.}
\affiliation{Istituto Nazionale di Ottica del Consiglio Nazionale delle Ricerche (CNR-INO), Largo Enrico Fermi 6, I-50125 Firenze, Italy.}

\author{Marco Pezzutto}
\affiliation{Centro de Física e Engenharia de Materiais Avançados (CeFEMA), Laboratory of Physics for Materials and Emerging Technologies (LaPMET), and PQI -- Portuguese Quantum Institute, Portugal.}
\affiliation{Istituto Nazionale di Ottica del Consiglio Nazionale delle Ricerche (CNR-INO), Largo Enrico Fermi 6, I-50125 Firenze, Italy.}

\author{Stefano Gherardini}
\affiliation{Istituto Nazionale di Ottica del Consiglio Nazionale delle Ricerche (CNR-INO), Largo Enrico Fermi 6, I-50125 Firenze, Italy.}
\affiliation{European Laboratory for Non-linear Spectroscopy, Università di Firenze, I-50019 Sesto Fiorentino, Italy.}

\begin{abstract}
We study the advantages of applying a sequential strategy for enhancing the precision in estimating the strength $p$ of a depolarizing channel, using qudit and two-qubit probes. The sequential strategy requires that the same depolarizing channel is repeatedly applied $M$ times to the quantum probe. From the analysis of the probe's output state, both the quantum Fisher information (QFI) and its regularization RQFI, given from multiplication by $p$, are analyzed. We determine that in the limit of $p \to 0$, the QFI is {\it linearly amplified} with the number of channel repetitions. Instead, in the weak-noise regime, the QFI is maximized at a finite value of $M$. Then, the RQFI decreases more quickly with $p$ as $M$ increases. In particular, for two-qubit probes, the RQFI decreases with the depolarization strength proportionally to the amount of correlation between the qubits. The greater the number of channel repetitions, the wider the range of $p$ for which the initial state of the probe that allows for the maximization of the QFI is a separable state. We also show that a single qubit probe undergoing a sequential encoding protocol outperforms the optimal two-qubit single-round QFI for $M>3$ in the weak-noise regime.  
\end{abstract}

\date{\today}

\maketitle

\section{Introduction}

Quantum technologies aim to perform tasks that are inaccessible with classical techniques, or to improve their performance beyond classical limits, by making use of dedicated quantum resources. The goal of quantum metrology is to determine the best precision at which an unknown parameter can be estimated~\cite{Giovannetti_2006, Matteo_G_Paris, PhysRevLett.96.010401, Giovannetti_2011}, even with significant advantages over classical methods~\cite{PhysRevLett.98.090401, PhysRevA.95.012111,Rafal2013}. The mathematical tool to achieve enhanced precision is the quantum Fisher information (QFI)~\cite{Carl_W_Helstrom, Christopher_M_Care, PhysRevA.85.022321, Jing_Liu}, which found successful application also in open quantum system scenarios~\cite{Qiang_Zheng, PhysRevA.94.042322, PhysRevA.94.042101}. Exploiting quantum squeezing and entanglement, quantum metrology of physical parameters~\cite{Luca_Pezze_2018, Zhihong_Ren_2019} is relevant, for instance, in quantum information and sensing~\cite{RevModPhys.89.035002, Pirandola_2018, Géza_Tóth}, for gravitational wave detection~\cite{Roman_Schnabel_2010}, quantum magnetometry~\cite{Brask_2015, Zhibo_Hou_2020}, atomic clocks~\cite{Augusto_2020}, and biology applications using nanodiamond quantum sensors~\cite{Taylor_2016, Yingke_Wu_and_Tanja_Weil}.

Quantum channels represent one of the fundamental building blocks of quantum information processing, and is reliably applied to describe open quantum dynamics~\cite{CarusoRMP2014}.
A common quantum approach for quantum-channel parameter estimation~\cite{M_Hotta} is the channel identification~\cite{Fujiwara_2001}, which is also referred to as quantum process tomography~\cite{nielsen2000quantum}. In such a context, it is worth determining the ultimate limits of parameter estimation in the presence of noise, and the best sensitivity for quantum noise characterization~\cite{Bylander2011, NorrisPRL2016, MullerSciRep2016, MullerSciRep2018}, where the parameter of interest is encoded through a quantum channel.

\begin{figure}[t]
    \centering
    \includegraphics[width=\columnwidth]{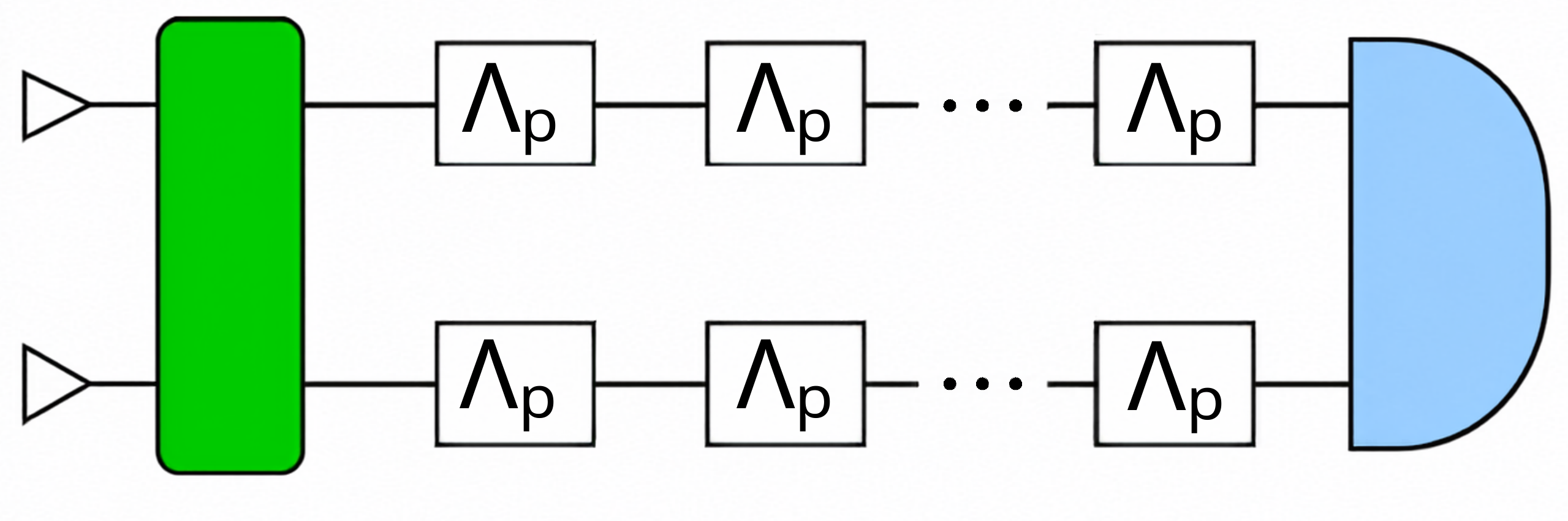} 
    \caption{Schematic illustration of the quantum metrology protocol considered in this work, for a two-qubit probe. The latter undergoes $M$ successive applications of the local depolarizing channel $\Lambda_p$. The examination of the probe's output state allows to determine the precision of estimating the depolarizing strength $p$ through the QFI. 
    }
    \label{fig:schematic}
\end{figure}

The depolarizing channel~\cite{nielsen2000quantum} is a standard model of quantum noise; it transforms any input state into a completely mixed state with a certain probability, effectively erasing the information contained in the input. The depolarizing channel has been widely studied in the contexts of quantum information processing and channel identification~\cite{1159773, PhysRevA.71.012322, Iyed_Ben_2007, Boixo_2008, Michael_2011}. Concerning the latter problem, a foundational contribution was given by Fujiwara~\cite{Fujiwara_2001}, who investigated the estimation problem of the isotropic depolarization parameter for a two-level quantum system, as an application of quantum entanglement~\cite{RevModPhys.81.865}.
Then, a joint optimization selecting the best input probe state and measurement strategy that minimizes the Bayes cost of a quadratic function was addressed in Ref.~\cite{sasaki_2002}. Later, in \cite{Boixo_2008}, an operational interpretation for pure-state global multipartite entanglement was provided in terms of the regularized QFI for the estimation of the strength of weak-noise local depolarizing channels. This reveled a direct link between the QFI associated with parameters entering a quantum channel and multipartite entanglement measures~\cite{PhysRevLett.87.270404, PhysRevA.85.022321}. In particular, it highlighted that entanglement can be employed to enhance precision of estimating parameters describing isotropic noise.

Recently, the problem of estimating the depolarizing-channel parameter has been investigated for the cases of qudit probes~\cite{Frey2011}, and qubit probes prepared in a generic mixed state~\cite{collins_2015}, comparing independent, sequential, and correlated-state protocols. In particular, in \cite{collins_2015} the Authors has shown that, for mixed initial states, correlated multi-qubit protocols can provide an estimation advantage over uncorrelated strategies in suitable parameter regimes. Moreover, it has been investigated how, in a sequence of quantum gates interspersed by depolarizing channels, the QFI of estimating the depolarization strength modifies with the number of gates; advantages over the case with a single depolarizing channel have been determined.

In this paper, starting from the results of Refs.~\cite{Frey2011,collins_2015}, 
we address the problem of determining the best precision in estimating the strength $p$ of a depolarizing channel, using qudit and two-qubit probes in a sequential metrology strategy. The latter relies on repeating $M$ times the depolarizing channel $\Lambda_p$ under scrutiny; see also Fig.~\ref{fig:schematic}. Examining how the output state of the probe depends on the noise parameter $p$ after repeated applications of the channel is the way to compute the QFI for the estimation of $p$. The maximum estimation precision is expressed in terms of the Cram\'er-Rao bound, and is achieved from maximizing the QFI, provided that the probe is initialized in a pure input state~\cite{Fujiwara_2001}. We thus determine analytically the behavior of the QFI and of its regularization RQFI, upon multiplication by $p$, as a function of both $p$ and the number of channel repetitions $M$. For the single-qubit case, our derivations are valid for any initial state of the probe, before the applications of the channel. Instead, for the case of two-qubit probes, we restrict our analysis to input states exhibiting the symmetric form $\sqrt{\mu} \ket{0}^{\otimes N} + \sqrt{1-\mu} \ket{1}^{\otimes N}$, where the parameter $\mu\in[0,1]$ characterizes the amount of correlations. 

The QFI associated with the sequential strategy is maximized by initializing the quantum probe in a pure input state. In the limit of $p \to 0$, the QFI is {\it linearly amplified} with the number $M$ of channel repetitions. This linear grows with $M$ is no longer guaranteed in the weak-noise regime ($p$ small but not infinitesimal), where the maximum of the QFI is achieved in correspondence of a finite value of $M$, which we denote as $M^*$. As $p$ increases, $M^*$ tends to $1$ according to a power law.

We compare our results with those of Ref.~\cite{Fujiwara_2001}, which investigates the optimal estimation of the depolarizing parameter $p$ using two qubits. The authors compare a parallel strategy involving a two-qubit probing system, in which the same quantum channel acts independently on each probe, with a single-qubit channel extension strategy~\cite{Fujiwara2008, Koodyski2013}, in which one qubit serves as the probe and the other as an auxiliary system. In the latter approach, the probe can be initially entangled with the auxiliary qubit, which does not undergo the encoding channel. The QFI is then optimized over joint probe-ancilla input states, yielding the so-called extended channel QFI. This quantity provides an upper bound on the achievable QFI and can guide the design of experimental protocols that, in some cases, achieve improved estimation precision~\cite{Wang2018}. 
Interestingly, here we find that the sequential strategy for the single-qubit probe gives rise to a sensitivity beyond the limit described in~\cite{Fujiwara_2001} in the weak-noise regime. This is not intended to be a strict comparison between the protocols, as they operate under different conditions and employ different resources. The statement above is intended as a useful guide to be taken into account when the repeated application of the depolarizing channel is not as costly as operating any of the forms of parallel strategy of metrology.

The enhancement of the sensitivity in estimating $p$ for $M>1$, with $p$ sufficiently small, is lost more rapidly as $p$ increases. In fact, it can be determined that the magnitude of the slope of the RQFI, evaluated with respect to $p$, linearly decreases with $M$ in the weak-noise regime. We perform this analysis, using analytical arguments, on two-qubit probes initialized in the state $\sqrt{\mu} \ket{0}^{\otimes N} + \sqrt{1-\mu} \ket{1}^{\otimes N}$.

We also investigate the role of quantum correlations (quantified by $\mu$) between the two qubits of the probing system, which undergoes the sequential strategy of metrology. We determine that the slope of RQFI as a function of $p$ in the weak-noise regime strictly depends on $\mu$: RQFI decreases with $\mu$ faster when $M$ increases. In particular, the optimal value of $\mu$, $\mu^*$, that maximizes the QFI is equal to $0$ or $1/2$ for any $M$ and $p$. $\mu=0$ corresponds to an initial separable state, while $\mu=1/2$ stands for the GHZ state. Interestingly, the greater the number of channel repetitions, the wider the range of $p$ for which the initial state of the probe that allows for the maximization of the QFI is a separable state.

The paper is organized as follows. Sec.~\ref{sec:depolarizing_strenght_estimation} introduces the framework for estimating the depolarizing strength through the sequential strategy, and defines the quantities used throughout the analysis. In Sec.~\ref{sec:single_probe}, we investigate the single probe scenario with a focus on the qubit case, and in Sec.~\ref{sec:extended} the sequential protocol is compared with the optimal single-round two-qubit strategy of Ref.~\cite{Fujiwara_2001}. In Sec.~\ref{sec:depol_derivation}, we extend the analysis to a two-qubit probe initialized in the state $\sqrt{\mu} \ket{0}^{\otimes N} + \sqrt{1-\mu} \ket{1}^{\otimes N}$, by  studying the role of quantum correlations and examining the behavior of the RQFI in the weak-noise regime. Finally, Sec.~\ref{sec:conclusion} summarizes the main conclusions and provides an outlook of our findings.

\section{Depolarizing strength estimation}
\label{sec:depolarizing_strenght_estimation}

As shown in Ref.~\cite{Boixo_2008}, the action of the single application of the depolarizing channel on the initial $N$-qudit state $\hat\rho$ is given by
\begin{equation}\label{eq:N_qudits}
\hat\rho_p \equiv \Lambda_p(\hat\rho)
= \left( 1 - p \sum_{j=1}^{N} d_j \right)\,\hat\rho
+ p \sum_{j=1}^{N}
\left(
\hat{\mathbb{I}}_j \otimes \mathrm{Tr}_j[\hat\rho]
\right),
\end{equation}
where $p$ is the depolarizing parameter we want to estimate, $\hat{\mathbb{I}}_j$ denotes the identity operator acting on the subsystem $j$ with dimension $d_j$, and $\mathrm{Tr}_j(\hat{\rho})$ is the partial trace over all the qudits except the $j$-th one. Eq.~(\ref{eq:N_qudits}) generalizes the action of a single-qudit depolarizing channel to multiple quantum probes, explicitly showing how depolarizing noise affecting each probe randomizes the global probes' state which induces a loss of purity. The depolarizing parameter satisfies $0\le p \le d/(d^2-1)$, which ensures that the channel remains completely positive and trace preserving.

The dynamical framework considered in this work is illustrated in Fig.~\ref{fig:schematic}, where a sequential encoding is considered: a quantum probe initialized in the state $\hat{\rho}$ is repeatedly ($M$ times) subjected to a depolarizing channel $\Lambda_{p}$, which encode the parameter $p$ into the final probes' state. We denote the $\hat{\rho}_{p}^{(M)}=\Lambda^{M}_p(\hat{\rho})$ the final state of the probes after $M$ applications of the depolarizing channel. The objective of quantum parameter estimation is to determine the ultimate precision achievable in estimating a parameter by measuring the state of the probes.
This precision is quantified through the QFI. In fact, the lower bound on the estimation uncertainty on $p$ is provided by the quantum Cram\'er-Rao bound~\cite{Helstrom1967}
\begin{equation}
    \Delta p \geq \frac{1}{\sqrt{\nu J\left(\hat\rho_{p}^{(M)}\right)}},
    \label{eq: QCRB}
\end{equation}
where $\nu$ denotes the number of independent measurements and $J (\hat{\rho}_p^{(M)})$ is the QFI associated with the estimation of the parameter $p$. The QFI can be evaluated with the formula
\begin{align}
    J\left(\hat\rho_{p}^{(M)}\right) = \Tr\left[ \left(\hat{L}_{p}^{(M)}\right)^{2}\hat\rho_{p}^{(M)} \right],
    \label{eq: QFI_rhop}
\end{align}
where $\hat{L}_{p}$ is the Symmetric Logarithmic Derivative (SLD), a $d^{N}\times d^N$ matrix implicitly defined by the Lyapunov equation
\begin{equation}
    \partial_p\hat\rho_p^{(M)} = \frac{1}{2}\left\{ \hat{L}_p^{(M)}, \hat\rho_p^{(M)} \right\}.
    \label{eq:lyapunov}
\end{equation}
In this work, Eq.~\eqref{eq:lyapunov} is solved directly to evaluate the QFI in certain cases, while alternative formulations are employed in others. In particular, the QFI can equivalently be expressed in terms of the spectral decomposition of the final state $\hat\rho_p^{(M)}$. Let $\{ \lambda_i \}$ and $\{|i\rangle\}$ be the eigenvalues and eigenvectors of $\hat\rho_p^{(M)}$, respectively. For $\lambda_i+\lambda_j \neq 0$, the entries $L_{i,j}$ of the SLD in the eigenbasis of $\hat\rho_p^{(M)}$ are 
\begin{align}
L_{i,j} = \frac{ 2\bra{i}\partial_p\hat\rho_p^{(M)}\ket{j} }{\lambda_i+\lambda_j},
\label{eq: L_ij}
\end{align}
while $L_{i,j}$ is set to zero if $\lambda_i+\lambda_j=0$. By substituting Eq.~\eqref{eq: L_ij} into Eq.\eqref{eq: QFI_rhop}, the QFI takes the form
\begin{align}
  J\left( \hat\rho_{p}^{(M)} \right)
  =\sum_{\substack{i,j\: \:\text{s.t.}\\ \lambda_i+\lambda_j>0}}
  \frac{2\,|\bra{i}\partial_p\hat\rho_p^{(M)}\ket{j}|^2}{\lambda_i+\lambda_j}.
  \label{eq:master}
\end{align}

Finally, we also introduce the Regularized Quantum Fisher Information (RQFI)~\cite{Boixo_2008} defined as
\begin{equation}
    \mathcal{J}\left( \hat\rho_p^{(M)} \right) = pJ\left( \hat\rho_p^{(M)} \right).
\label{eq:RQFI}
\end{equation}
The multiplication by the depolarizing strength $p$ regularizes the weak-noise divergence of the QFI discussed in the following. Hence, this quantity provides a finite figure of merit for comparing the estimation performance under different system dimensions and channel repetitions.

\section{Single quantum probe}
\label{sec:single_probe}

\begin{figure*}[ht]
\centering
\begin{subfigure}[b]{0.49\textwidth}
    \centering
    \includegraphics[width=\linewidth]{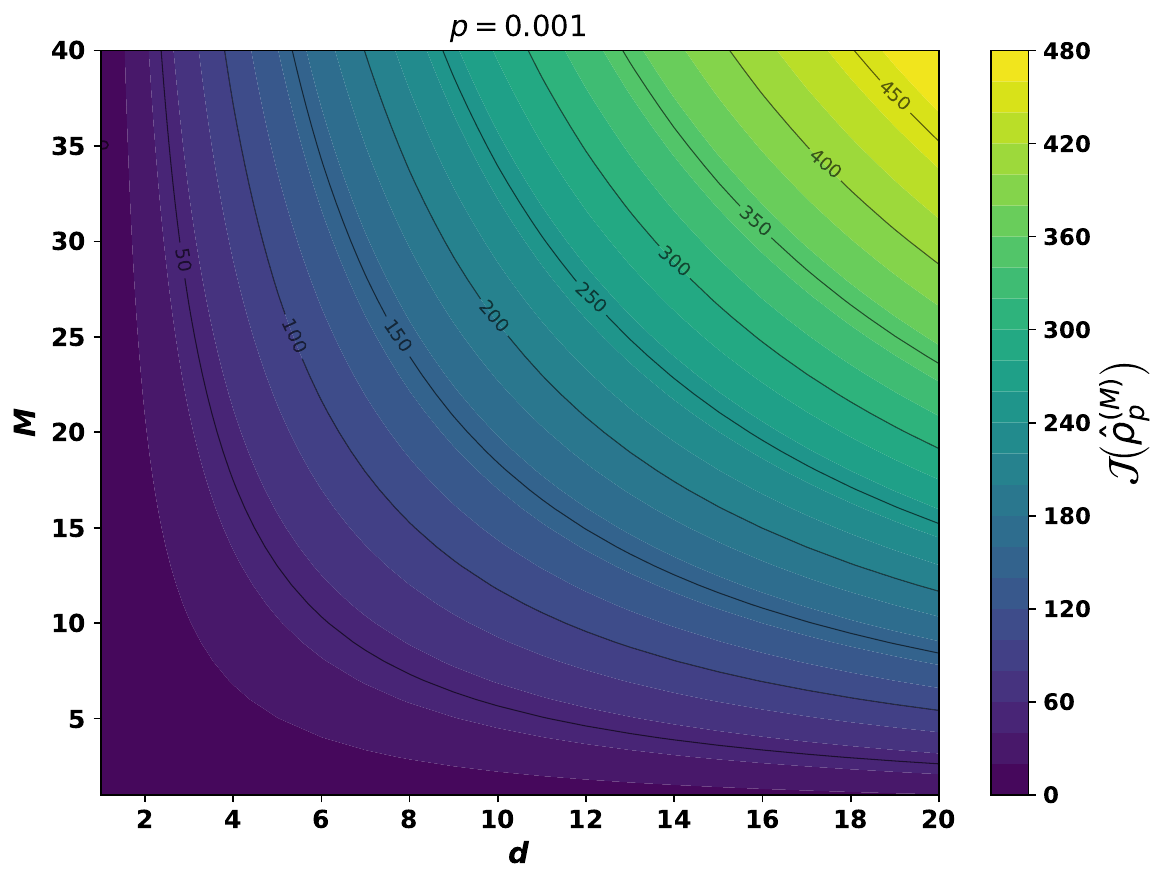}
    \caption{$p=0.001$}
    \label{GMGd_a}
\end{subfigure}
\hfill
\begin{subfigure}[b]{0.49\textwidth}
    \centering
    \includegraphics[width=\linewidth]{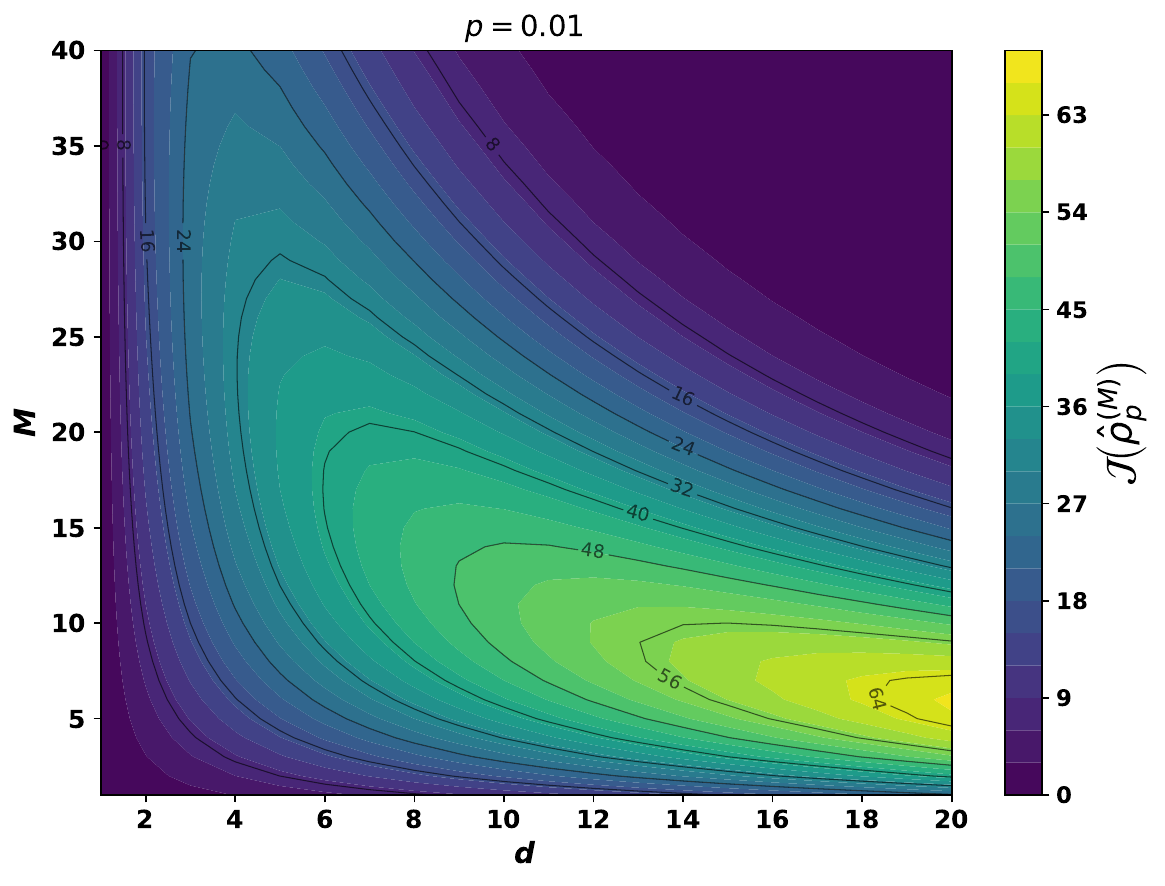}
    \caption{$p=0.01$}
    \label{fig:GMGd_b}
\end{subfigure}
\caption{RQFI as a function of the probe's dimension $d \in [2,20]$ and of the number $M \in [1,40]$ of channel applications. Panel (a): $p=0.001$. The RQFI is maximized by the highest-dimensional probe subjected to the maximum number of channel applications. Increasing both $d$ and $M$ enhances the precision of the metrological system in agreement with Eq.~\eqref{eq:qfi_small_epsilon_generic_M}. Panel (b): $p=0.01$. The interplay between $d$ and $M$ becomes less trivial. In fact, the best metrological performance are achieved by the highest dimensional system $d=20$ with $M=6$ repetitions. Unlike panel (a), for $d=20$, applying the depolarizing channel to the probe more times than $M=6$ decreases the RQFI.}
\label{fig:Generic_d_Generic_M}
\end{figure*}

In this section, we derive the QFI of a single ($N=1$) $d$-dimensional quantum probe undergoing multiple applications of the depolarizing channel $\Lambda_p$. We determine the exact value of the QFI for any $p$, and we compare it with the result in \cite{Boixo_2008} thus showing the importance of the higher order corrections terms. All the detailed derivation steps are included in the Appendix \ref{sec:SQP}. 

Let the initial state $\hat{\rho}$ of the probe be a pure state, and let us define the projector $\hat{Q}=\hat\id-\hat\rho$, such that $\hat\rho\hat Q=\hat Q\hat\rho=0$. Upon defining 
\begin{equation}
\xi_d(p) := 1-dp, 
\end{equation}
the evolution of a qudit under a generic number $M$ of application the depolarizing channel is
\begin{equation}
    \hat\rho^{(M)}_{p} = \left(\xi_d(p)^{M}+\frac{1-\xi_d(p)^{M}}{d}\right)\hat\rho
    +\frac{1-\xi_d(p)^{M}}{d}\hat Q.
    \label{eq:GDGMsingleprobe}
\end{equation}
We parametrize the SLD operator as a linear combination of the initial state $\hat\rho$ and the projector $\hat{Q}$, obtaining 
\begin{align}
    \hat L^{(M)}_{p} = a_{M}(p)\hat\rho+b_{M}(p) \hat Q,
    \label{eq:sld_singleprobe_genericM}
\end{align}
where 
\begin{eqnarray}
    a_{M}(p)&=&-\frac{Md(d-1)\xi_d(p)^{M-1}}{1+(d-1)\xi_d(p)^M},\\
    b_{M}(p)&=&\frac{Md\xi_d(p)^{M-1}}{1-\xi_d(p)^M}.
    \label{eq: AMBM}
\end{eqnarray}
The parametrization in Eq.~\eqref{eq:sld_singleprobe_genericM} for $\hat{L}_{p}^{(M)}$ is motivated by the structure of the output state $\hat\rho_{p}^{(M)}$, which can be written as a linear combination of the orthogonal projectors $\hat\rho$ and $\hat{Q}$. Since both $\hat\rho_{p}$ and its derivative $\partial_{p}\hat{\rho}_{p}$ lie within the operator subspace spanned by $\hat\rho$ and $ \hat{Q}$, the SLD $\hat{L}_{p}^{(M)}$, defined through Eq.~\eqref{eq:lyapunov}, must belong to the same subspace. Therefore, expressing $\hat{L}_{p}^{(M)}$ as a linear combination of $\hat\rho$ and $\hat{Q}$ suffices, reducing the Lyapunov equation to scalar relations that allow for direct determination of the coefficients $a_M(p)$ and $b_M(p)$.

Now, by substituting Eqs.~\eqref{eq:GDGMsingleprobe} and \eqref{eq:sld_singleprobe_genericM} into Eq.~\eqref{eq: QFI_rhop}, we obtain 
\begin{equation}\label{eq:MA}
    J\left( \hat\rho_{p}^{(M)} \right)
    = \frac{d^2(d-1)M^{2}\xi_d(p)^{2(M-1)}}
    {\left(1+(d-1)\xi_d(p)^{M}\right)\left( 1-\xi_d(p)^{M} \right)}.
\end{equation}
Eq.~\eqref{eq:MA} holds for generic depolarizing strength $p$ and generic number $M$ of channel applications, and represents the first result of the paper. 

Let us now discuss how Eq.~\eqref{eq:MA} simplifies in the weak-noise regime ($p$ close to zero), which is the case analyzed in \cite{Boixo_2008}. Expanding the coefficients in powers of $p$ yields
\begin{align}
J\left( \hat{\rho}_{p}^{(M)} \right)
&= \frac{M(d-1)}{p}
+\frac{M(d-1)}{12}
\Bigg[
6\!\left(3d-M(d+2)\right)
\notag\\
&
+\Big(
d^{2}(M^{2}-12M+23)
-24Md
+12M^{2}
\Big)p
\Bigg]
\notag\\
& + O(p^{2}),
\label{eq:qfi_small_epsilon_generic_M}
\end{align}
from which, by substituting $M=1$, we obtain a direct extension of formulas in \cite{Boixo_2008} for a qudit probe up to the first order in $p$:
\begin{equation}
J(\hat\rho_{p})=\frac{d-1}{p}+(d-1)^2+(d-1)^3p+O(p^2).
\label{eq:qfi_sq_m1_smallp}
\end{equation}
Eqs.~\eqref{eq:qfi_small_epsilon_generic_M} and \eqref{eq:qfi_sq_m1_smallp} exhibit the characteristic $1/p$ divergence of the QFI for vanishing $p$, indicating that the depolarizing strength $p$ can be estimated with increasingly high precision as $p\to 0$. The finite order corrections improve the approximation of the QFI in the weak-noise regime.

\begin{figure}[t]
    \centering
    \includegraphics[width=0.975\columnwidth]{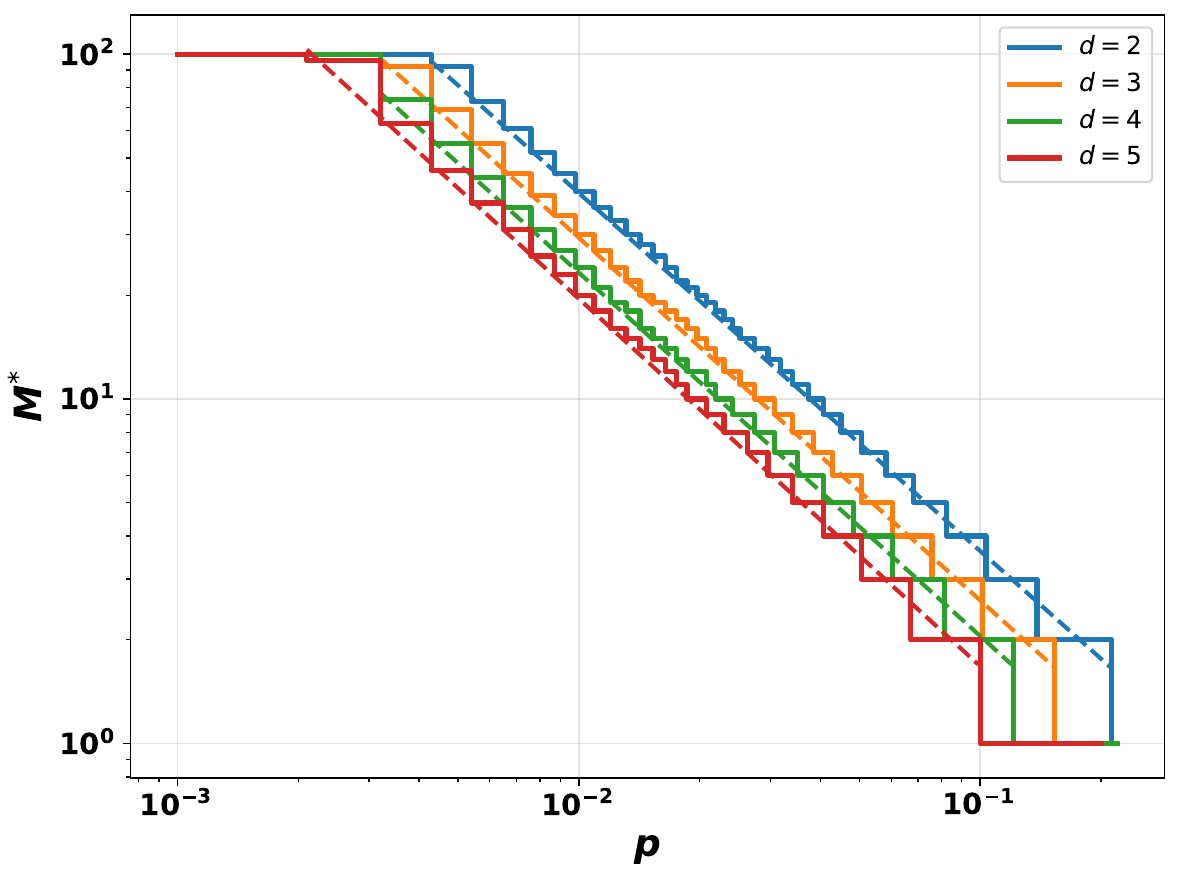} 
    \caption{
    Optimal number of channel applications, $M^*$, as a function of the depolarizing parameter $p$ for single probes with dimension $d=2,3,4,5$. The logarithmic scale on both axes reveals that, as long as $p$ is sufficiently small, the scaling behavior of the optimal probing strategy is well approximated by a $\sim 1/p$ power law, as demonstrated by the linear fits (dashed lines). As the noise strength increases, $M^*$ decreases rapidly, whereas increasing $d$ reduces the number of channel applications required to attain the maximum estimation precision.
    }
    \label{fig:Optimal_M_function_of_p}
\end{figure}

It is instructive to examine the interplay between the dimension $d$ of the probe and the number $M$ of channel applications, and its dependence on the parameter of interest. In Fig.~\ref{fig:Generic_d_Generic_M} the RQFI is plotted as a function of $d$ and $M$, for (a) $p=0.001$ and (b) $p=0.01$. For the smallest value of $p$, an increase in the RQFI is observed with both $M$ and $d$. This is consistent with Eq.~\eqref{eq:qfi_small_epsilon_generic_M}, which predicts $\mathcal{J}( \hat\rho_p^{(M)})\approx M(d-1)$ in the weak-noise regime. Conversely, increasing $p$ leads to a more intricate dependence of the QFI on $d$ and $M$. In particular, for $p=0.01$, the highest QFI is attained for a relatively small number of applications, $M=6$. Moreover, for certain values of $M$ (e.g., $M>15$), increasing the probe dimension can reduce the estimation precision. In Fig.~\ref{fig:Optimal_M_function_of_p} we plot $M^*$, the optimal value of $M$ such that the QFI is maximized, as a function of $p$, for different values of $d$. For $p$ sufficiently small, the observed trend of $M^*$ is well approximated by a $\sim 1/p$ power-law decay, as demonstrated by the fits. In addition, for $d=2$, starting from Eq.~\eqref{eq:MA} and treating $M$ as a continuous variable, it is possible to obtain the following analytic expression for $M^*$ as a function of $p$:
\begin{equation}
M^*(p) = \frac{-W\left(-\frac{2}{e^2}\right)-2}{2 \log (1-2 p)},
\label{eq:Mstar}
\end{equation}
where $W(z)$ is the Lambert W function with argument $z$. Assuming the positivity of the argument of the logarithm at the denominator, Eq.~\eqref{eq:Mstar} can be only applied in the range $p \in [0,1/2)$. 
For small $p$, by taking the power series expansion of Eq.~\eqref{eq:Mstar} around $p=0$, one recovers the $1/p$ behavior. 

\subsection{Qubit case}
\label{sec:qubit}

In this Section, we specify the analysis of the QFI above for a single-qubit probe ($d=2$) undergoing the sequential metrology protocol of Fig.~\ref{fig:schematic}.
The state of the qubit probe after $M$ applications of the depolarizing channel is straightforwardly determined by substituting $d=2$ in Eq.~\eqref{eq:GDGMsingleprobe}:
\begin{equation}
    \hat\rho^{(M)}_p=\xi_2(p) ^{M}\hat\rho+\frac{1-\xi_2(p) ^{M}}{2}\hat\id.
\end{equation}
The expression of the QFI can be similarly obtained from Eq.~\eqref{eq:MA}, whereby one has 
\begin{equation}\label{eq:J_M_qubit}
    J\left( \hat\rho_{p}^{(M)} \right) = \frac{4 M^{2}\xi_2(p)^{2(M-1)}}
    {\left(1+\xi_2(p)^{M}\right)\left( 1-\xi_2(p)^{M} \right)}.
\end{equation}
Moreover, in the weak-noise regime, Eq.~\eqref{eq:J_M_qubit} takes the simplified expression
\begin{equation}
    J\left( \hat\rho_{p}^{(M)} \right)
    = \frac{M}{p} + M(3-2M) + O(p).
    \label{eq:J_M_qubit_small_p}
\end{equation}
The leading-order term, $M/p$, indicates that each application of the depolarizing channel $\Lambda_p$ contributes an additive amount of information about the noise parameter $p$, resulting in a QFI that grows linearly with $M$. Thus, in the $p \to 0$ limit, the sequential protocol of Fig.~\ref{fig:schematic} provides a metrological advantage that grows with the number of channel applications, without requiring additional ancillary qubits.

\begin{figure}[t]
    \centering
    \includegraphics[width=0.975\columnwidth]{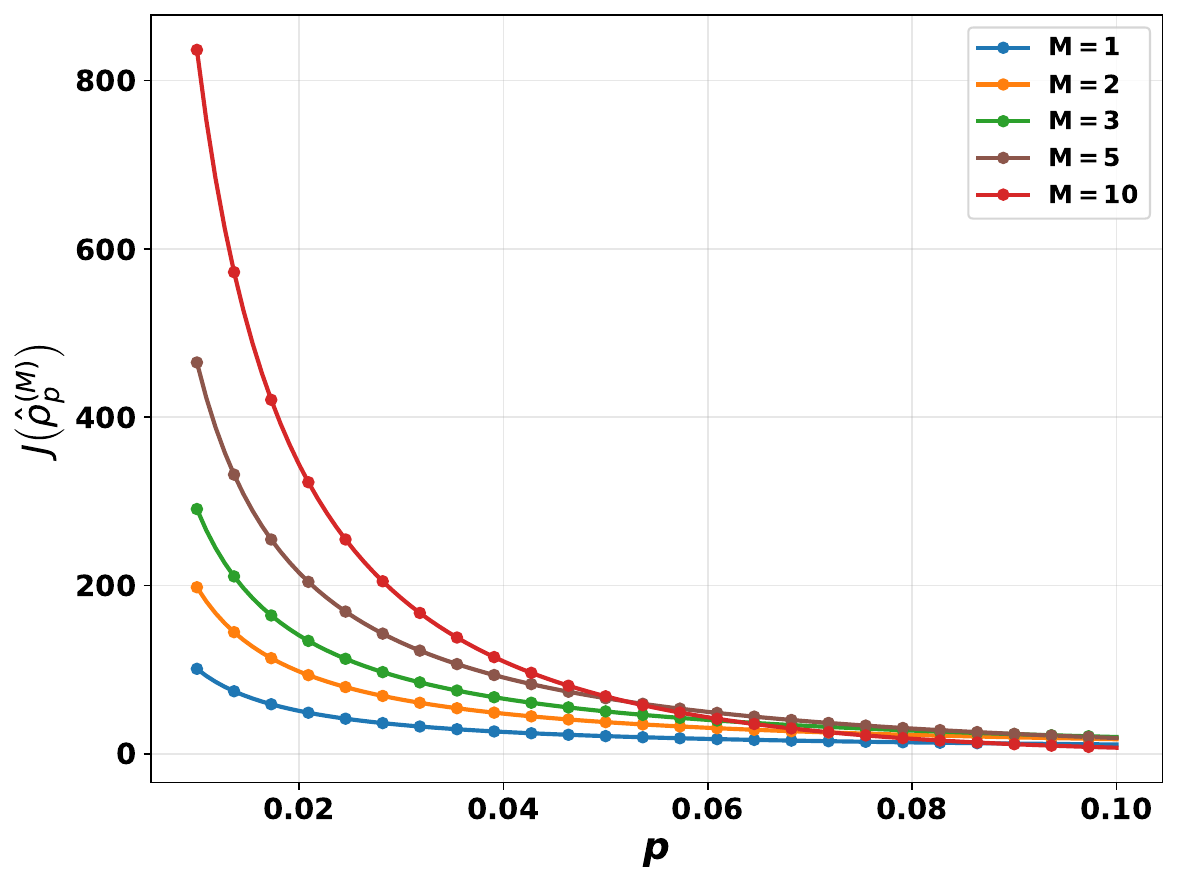} 
    \caption{QFI as a function of the depolarizing strength $p$ for a single-qubit probe initially prepared in the pure state $|0\rangle\!\langle 0|$, with $M=1,2,3,5,10$. The numerical results (markers) are in perfect agreement with the corresponding analytical expressions (solid lines). Larger values of $M$ enhance the QFI in the weak-noise regime. For every $M$, increasing $p$ suppresses the estimation precision.
    }
    \label{fig:QFI_vs_p_comparison}
\end{figure}

In Fig.~\ref{fig:QFI_vs_p_comparison} we show the QFI as a function of the depolarizing strength $p$ for a single-qubit probe initially prepared in the pure state $|0\rangle\langle0|$, computed using Eq.~\eqref{eq:J_M_qubit}, with $M=1,2,3,5,10$. The solid lines represent the analytical expressions, while the markers correspond to numerical results. The exact overlap of the markers with the solid curves for all $M$ confirms the agreement between the analytical and numerical results. Consistently with Eq.~\eqref{eq:J_M_qubit_small_p}, the QFI increases significantly with $M$ in the weak-noise regime, confirming that the sequential strategy amplifies the probe's sensitivity to the channel parameter in this regime. Then, the QFI decreases monotonically with $p$ for any $M$. Hence, as $p$ increases, the probe's state becomes increasingly insensitive to variations in $p$ for any $M$, so estimating the depolarizing strength precisely becomes harder. 

While the discussion presented above assumes a pure initial state of the probe, an analytical expression of the QFI for a generic initial state can be derived when the probe is a single qubit. This makes use of the Bloch representation, in which any generic single-qubit initial state can be written as 
\begin{equation}
    \hat\rho = \frac{1}{2}\left( \hat\id + \boldsymbol{w}\cdot\boldsymbol{\sigma} \right),
    \label{eq:initial_state_bloch}
\end{equation}
where $\boldsymbol{\sigma} = (\hat\sigma_x,\hat\sigma_y,\hat\sigma_z)$ is the vector of the Pauli matrices and $\boldsymbol{w} = (w_x, w_y, w_z)$ such that $\abs{\boldsymbol{w}}\leq 1$ denotes the Bloch vector comprising the expectation values of the Pauli matrices with respect $\hat{\rho}$. 
The purity of the state depends on the modulus of the Bloch vector as $\text{Tr}[\hat{\rho}^2] = (1+\abs{\boldsymbol{w}}^2)/2$. 

Within this formalism, the dynamics is described by the evolution of the Bloch vector. Specifically, the application of $M$ depolarizing channels evolves the state as
\begin{equation}
    \hat\rho_p^{(M)} = \frac{1}{2}\left( \hat\id + \boldsymbol{w}_p^{(M)}\cdot\boldsymbol{\sigma} \right),
    \quad
    \boldsymbol{w}_p^{(M)} = \xi_2(p)^M \boldsymbol{w},
    \label{eq:final_state_bloch}
\end{equation}
where the final Bloch vector is isotropically contracted by a factor $\left| \xi_2(p) \right| =\left| 1-2p \right|^M<1$ and the purity is reduced accordingly. 
In this representation, the QFI of the final state is given by~\cite{Zhong2013quantum} 
\begin{equation}
J\left(\hat\rho_{p}^{(M)}\right) =
\left| \partial_p \boldsymbol{w}_p^{(M)} \right|^2
+ \dfrac{
\left(\boldsymbol{w}_p^{(M)} \cdot \partial_p \boldsymbol{w}_p^{(M)} \right)^2
}
{
1-\left|\boldsymbol{w}_p^{(M)}\right|^2
}.
\end{equation}
Then, Eq.~\eqref{eq:final_state_bloch} leads to
\begin{equation}
    J\left(\hat\rho_{p}^{(M)}\right) = \frac{4M^2\xi_2(p)^{2(M-1)}\abs{\boldsymbol{w}}^2}{1-\xi_2(p)^{2M}\abs{\boldsymbol{w}}^2}.
\label{eq:final_qfi_sq}
\end{equation}
This result was also obtained in~\cite{collins_2015} using a different parameterization. One can observe that only the length of the final Bloch vector affect the QFI in Eq.~\eqref{eq:final_qfi_sq}, as the depolarizing channel acts without selecting a preferred eigenbasis. Due to Eq.~\eqref{eq:final_state_bloch}, the QFI depends on the initial state only through its purity.

\begin{figure}[t]
    \centering
    \includegraphics[width=0.975\columnwidth]{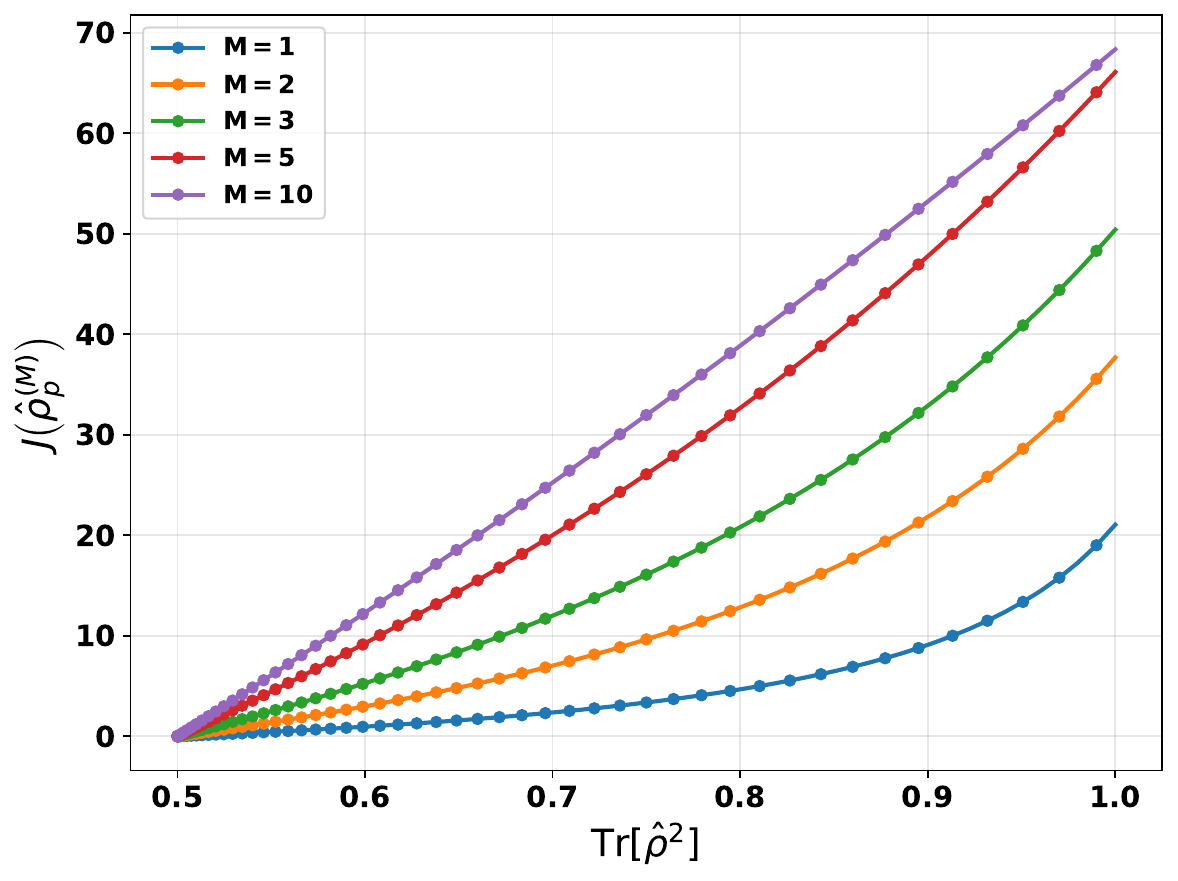} 
    \caption{QFI as a function of the initial state purity $\mathrm{Tr}[\hat{\rho}^2]$ for a single-qubit probe undergoing repeated depolarizing channel applications, with $M=1,2,3,5,10$ and $p=0.05$. The probe is initialized in a state with equal populations $(w_z = 0)$, while the coherence magnitude is varied to tune the purity from the maximally mixed state $\mathrm{Tr}[\hat{\rho}^2]=1/2$ to the pure state limit $\mathrm{Tr}[\hat \rho^2]=1$. The numerical results are shown by markers connected with dashed lines, whereas the analytical predictions are represented by solid lines.}
    \label{fig: QFI vs p for a Generic state}
\end{figure}

For this reason, in the numerical evaluations we perform we set $w_y = w_z = 0$ without loss of generality, and we tune the purity of the initial state through the value of $w_x \in [0,1]$. 
Fig.~\ref{fig: QFI vs p for a Generic state} shows the QFI as a function of the initial state purity $\text{Tr}[\hat{\rho}^2]$ for a single-qubit probe undergoing repeated $M=1,2,3,5$ and $10$ applications of the depolarizing channel, with $p=0.05$. The analytical results (solid lines) perfectly agree with the numerical results (markers). As purity increases, the the QFI increases monotonically for all values of $M$ and $p$. This behavior suggests that purer probe initial states are more sensitive to changes in the depolarizing parameter, resulting in better estimation precision. In contrast, the maximally mixed state is insensitive to the depolarizing parameter because it remains unchanged under the action of the depolarizing channel. Since the output state is independent of $p$ in this case, it follows that the corresponding QFI is zero. The observed monotonic increase in the QFI with the initial state purity generalizes the results in \cite{Fujiwara_2001} for $M=1$, establishing that pure probe states are optimal for estimating the depolarizing parameter.

\subsection{RQFI and comparison between single-qubit sequential vs.~two-qubit single-round strategy}
\label{sec:extended}

\begin{figure}[t]
    \centering
    \includegraphics[width=0.975\columnwidth]{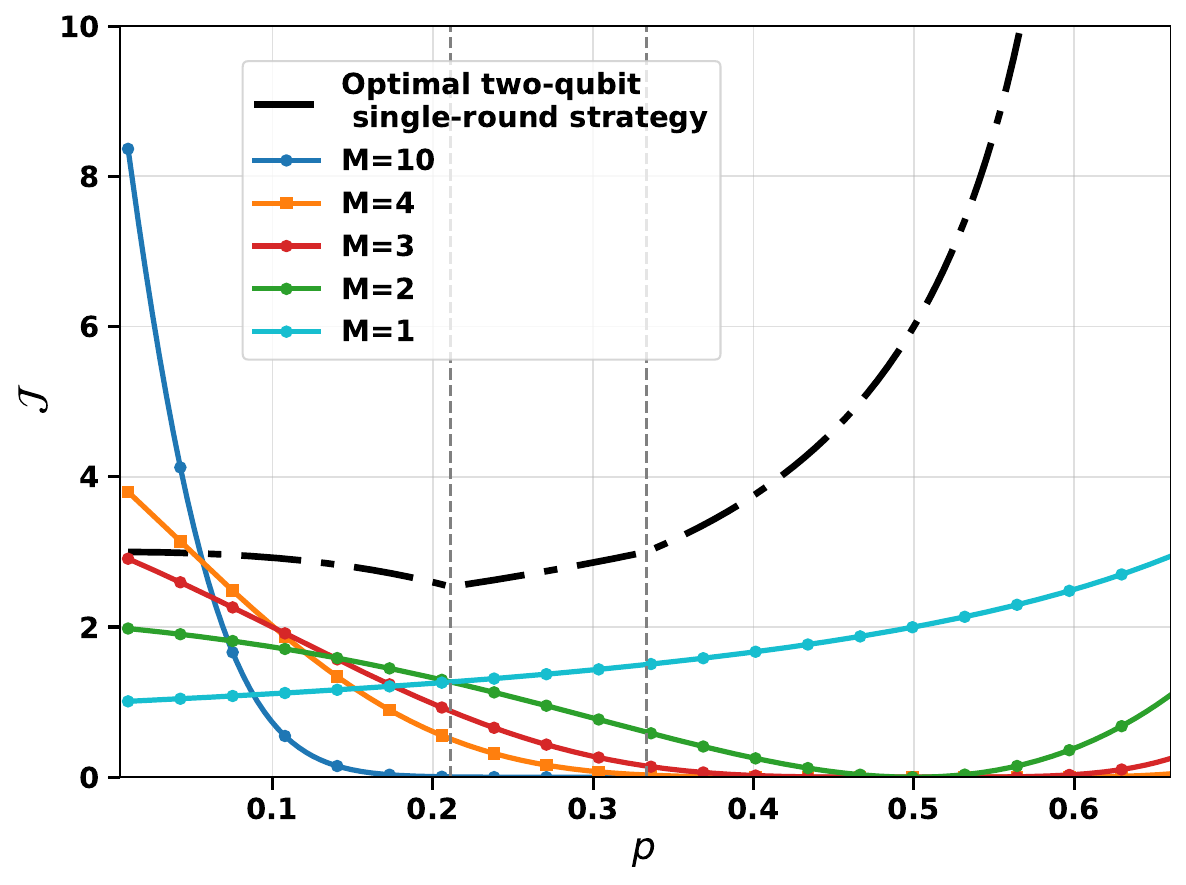}    
    \caption{Comparison between the RQFI of the optimal two-qubit single-round strategy described in~\cite{Fujiwara_2001} (the transition points of this strategy described in the main text are marked by vertical dashed lines) and the single-qubit sequential strategy with pure initial state for $M=1, 2, 4, 10$. While the optimal single-round strategy outperforms the second for $M \leq 3$, the sequential strategy performs better for larger values of $M$ in the weak-noise regime.
    }
\label{fig:qfi_comparison}
\end{figure}

The RQFI provides further insight on the advantage of the sequential strategy, by removing the leading $1/p$ divergence of the QFI in the $p\to 0$ limit. The corresponding curves for different values of $M$ are shown as solid lines in Fig.~\ref{fig:qfi_comparison}.
As the RQFI is obtained from the QFI rescaled by $p$, these two quantities have the same ordering across the curves corresponding to different $M$, and crossing points placed at the same values of $p$. For $M=1$, the RQFI increases with $p$, reaches $\mathcal{J} = 2$ at $p=1/2$, whereas for $M>1$ it decreases monotonically and approaches zero as the $p$ increases. This behavior indicates that the sequential approach offers limited metrological sensitivity at large depolarization strengths.

We now investigate under what conditions repeated applications of the depolarizing channel on a single-qubit probe can yield an advantage over the best performance of single-round metrology of $p$ with a two-qubit probe. 
The optimal strategy for estimating $p$ using two qubits in a single encoding round was identified in Ref.~\cite{Fujiwara_2001}. It involves three different approaches, depending on the value of $p$, namely applying $\Lambda_p\otimes\Lambda_p$ on the maximally entangled input state ($\mu=1/2$) for $0\leq p\leq(1-1/\sqrt{3})/2$; still applying $\Lambda_p\otimes\Lambda_p$ but on the pure separable input state ($\mu=0,1)$ for $(1-1/\sqrt{3})/2<p<1/3$; and implementing $\Lambda_p\otimes \id$ on the maximally entangled input state ($\mu=1/2$) for $1/3\leq p\leq2/3$, where $\id$ here denotes the identity channel.
The RQFI corresponding to this strategy is plotted as a black dot-dashed black line in Fig.~\ref{fig:qfi_comparison}.

The QFI obtained by considering $\Lambda_p \otimes \id$ is commonly referred to as the extended QFI, and provides an upper bound on the QFI achievable with a single-qubit probe subjected to one application of the encoding channel. This bound cannot be saturated for the depolarizing strength estimation task~\cite{Koodyski2013}. Hence, as the optimal two-qubit single-round strategy is equal or greater than the extended QFI, such a strategy provides an upper bound on the channel's RQFI for a single-qubit probe achieved by a single application of $\Lambda_p$, yielding values greater than those of the $M=1$ curve. Even the curve for $M=2$ lies below that of the optimal two-qubit single-round strategy, showing that the latter achieves better estimation performance for the {\it same} total number of channel applications. In experimental photonic settings, repeating a depolarizing channel is as feasible as applying the channel once~\cite{wang2025}, meaning that the sequential strategy can be more convenient than using entangled two-qubit probes. This motivates comparing the optimal two-qubit single-round strategy with the RQFI obtained by applying the channel sequentially $M>2$ times to a single-qubit probe.
We observe that increasing the number of applications to $M\geq 4$ allows the sequential strategy to outperform the optimal single-round one of~\cite{Fujiwara_2001} in the weak-noise regime. Although more channel applications are used---which inevitably requires a longer experiment duration---our results show that a sequential strategy compensate for or even surpass the performance given by the addition of a qubit to the experimental setup in the weak-noise regime.

\section{Two-qubit probe}
\label{sec:depol_derivation}

We extend our analysis to a two-qubit system, prepared in a pure possibly correlated quantum state, that undergoes $M$ repeated applications of the depolarizing channel. The corresponding final state is described by
\begin{align}
    \hat{\rho}_p^{(M)}=& \:\xi_2(p)^{2M}\hat\rho+\frac{\xi_2(p)^{M}-\xi_2(p)^{2M}}{2} S(\hat\rho)\nonumber\\
    &+\frac{1-2\xi_2(p)^{M}+\xi_2(p)^{2M}}{4}\hat{\id}_{4},
    \label{eq:Depol Chan}
\end{align}
where $S(\hat\rho):=\hat\id_2\otimes\Tr_{Q_1}[\hat\rho] + \Tr_{Q_2}[\hat\rho]\otimes\hat\id_2$, with $\Tr_{Q_i}[\hat\rho]$ denoting the partial trace over the $i$-th qubit.
The detailed derivation is given in Appendix~\ref{sec:two_qubit_depol_derivation}.
Eq.~\eqref{eq:Depol Chan} holds for a generic pure input state whose qubits have identical reduced states, i.e., $\Tr_{Q_2}[\hat\rho] = \Tr_{Q_1}[\hat\rho]$.
The first term in the right-hand side of Eq.~\eqref{eq:Depol Chan} preserves the original structure of the initial state, while the remaining terms progressively suppress its coherence and correlations as the channel is applied repeatedly. Repeated applications of the depolarizing channel drive the probes away from their original state, towards the maximally mixed state following an exponential trend. As a result, the quantum probes gradually lose their ability to encode information about the parameter $p$ as $M$ increases.

In the weak-noise regime, the QFI for $M$ channel applications can be obtained in closed form by expanding Eq.~\eqref{eq:Depol Chan} to leading orders in $p$, for a generic pure initial state. The full derivation is given in Appendix~\ref{sec:app_two_qubit_weak_noise}. 
In particular, the leading order QFI is
\begin{equation}
J\left(\hat\rho_{p}^{(M)}\right) = \frac{M(4-2P)}{p} + O(1),
\label{eq:JM_main}
\end{equation}
where $P := \Tr_{Q_2} [\hat{\rho}^2] = \Tr_{Q_1} [\hat{\rho}^2]$. The QFI \eqref{eq:JM_main} diverges as $p\to0$ and the RQFI scales linearly with the number of channel applications: 
\begin{equation}
    \mathcal{J}( \hat\rho_p^{(M)})=M\mathcal{J}( \hat\rho_p^{(1)})+O(1).
\end{equation}
For a fixed $M$, the RQFI decreases as the purity $P$ of the reduced single-qubit state increases, i.e., as the two-qubit state becomes less entangled.
In fact, the coefficient multiplying the $M/p$ term can be written as $4-2P=2+\varepsilon(\ket{\psi})$~\cite{Boixo_2008}, where $\varepsilon(\ket{\psi}) = 2[1-1/2( P_{Q_1} + P_{Q_1})] = 2(1 - P)$ is the Meyer--Wallach entanglement measure~\cite{Meyer2002}, equal to the average linear entropy of the single-qubit reduced states~\cite{brennen2003observablemeasureentanglementpure}. This result indicates an enhanced metrological performance provided by entangled initial states in the weak-noise regime.

\begin{figure}[t]
    \centering
    \includegraphics[width=0.975\columnwidth]{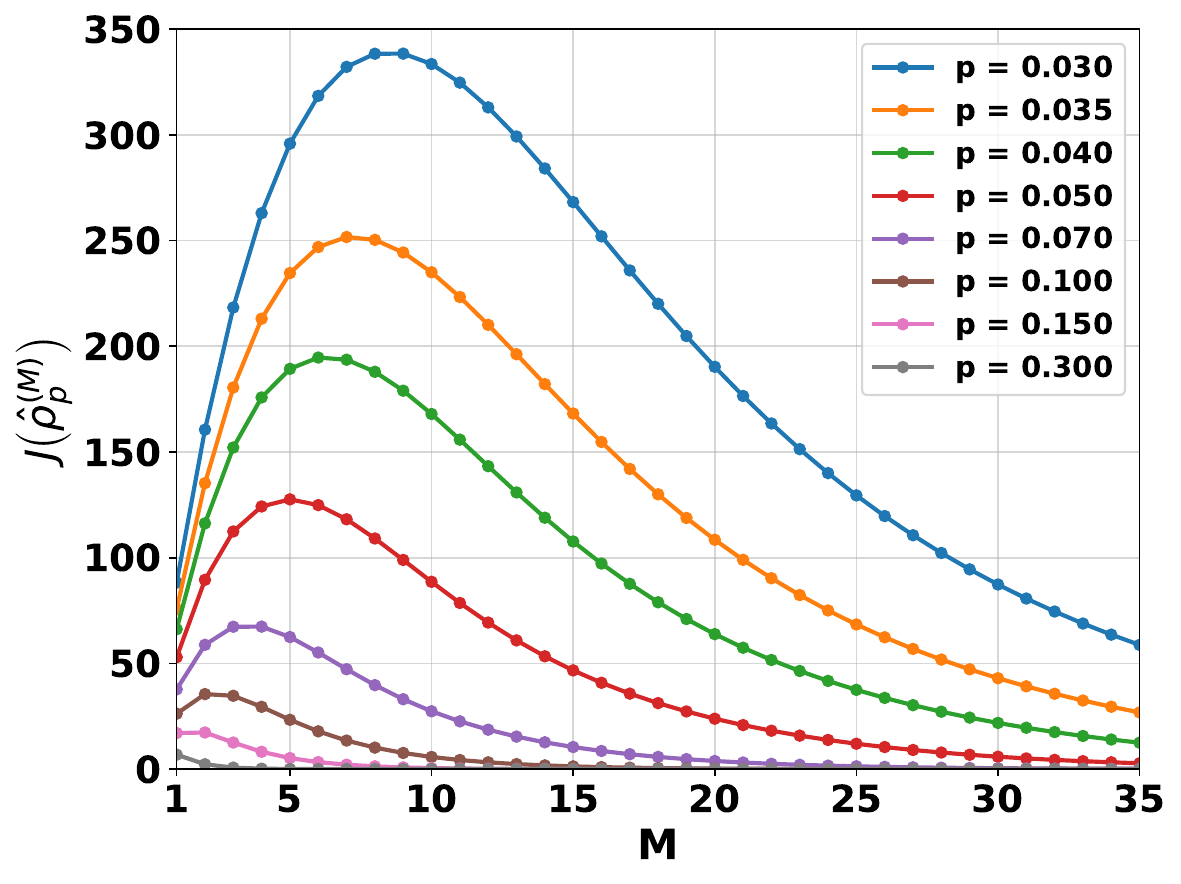} 
    \caption{QFI as a function of $M$ for different values of the depolarizing strength $p$ for two-qubit probes prepared in the state of Eq.~\eqref{eq: initial state} with $\mu = 0.2$. In the weak-noise regime, the QFI initially increases with $M$, reaches a maximum, and then decreases when repeated applications suppress the information encoded into the probes' state. As $p$ increases, both the maximum attainable QFI and the optimal number of channel applications decrease. The numerical results are shown by markers connected with dashed lines, whereas the analytical predictions are represented by solid lines.}
    \label{fig: QFI vs M}
\end{figure}

Let us now consider, as initial state of the two-qubit probe, the pure state
\begin{align}
|\psi\rangle = \sqrt{\mu}|00\rangle+\sqrt{1-\mu}|11\rangle, 
\label{eq: initial state}
\end{align}
where the parameter $\mu\in[0,1]$ tunes the initial correlations: $\mu=1/2$ corresponds to a maximally entangled state and $\mu=0,1$ to a separable product state. In Fig.~\ref{fig: QFI vs M} we show the QFI as a function of $M$, for different values of the depolarizing strength $p$, for a two-qubit probe initialized in the state of Eq.~\eqref{eq: initial state} with $\mu=0.2$. The expression of this QFI is analytical but quite intricated; for this reason, we report it in Appendix~\ref{app:qfi_for_specific_two_qubit_initial_state}.
The QFI exhibits non-monotonic behaviour:  it initially increases with $M$, reaches a maximum at some optimal number of applications $M^*$ and subsequently decreases for larger $M$. The peak of this maximum depends strongly on $p$. For a small value of $p$, e.g. $p=0.035$, the QFI reaches a high peak value at $M^*=7$, reflecting the fact that the probe can undergo multiple channel applications at fixed $p$ before decoherence becomes significant in the weak-noise regime. As $p$ increases, the peak value of the QFI decreases substantially and shifts towards smaller $M^*$.  For the largest values of $p$ considered ($p = 0.15$ and $0.3$), the QFI is maximal at $M^* = 1$ and decreases monotonically with larger $M$. This means that repeated applications of the channel provide no metrological advantage in the strong-noise regime.     

\begin{figure*}[t]
\centering
\begin{subfigure}[b]{0.32\textwidth}
    \centering
    \includegraphics[width=\linewidth]{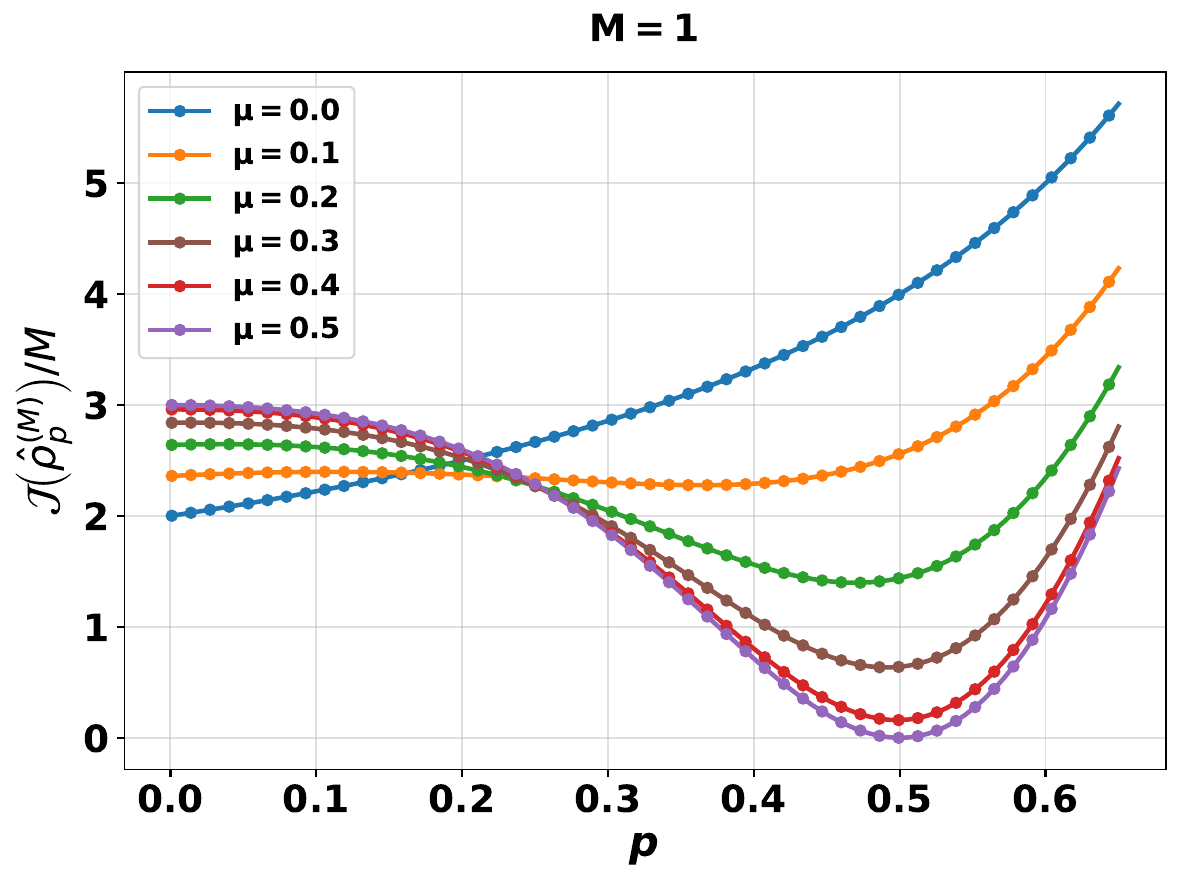}
    \caption{}
    \label{fig:1a}
\end{subfigure}
\hfill
\begin{subfigure}[b]{0.32\textwidth}
    \centering
    \includegraphics[width=\linewidth]{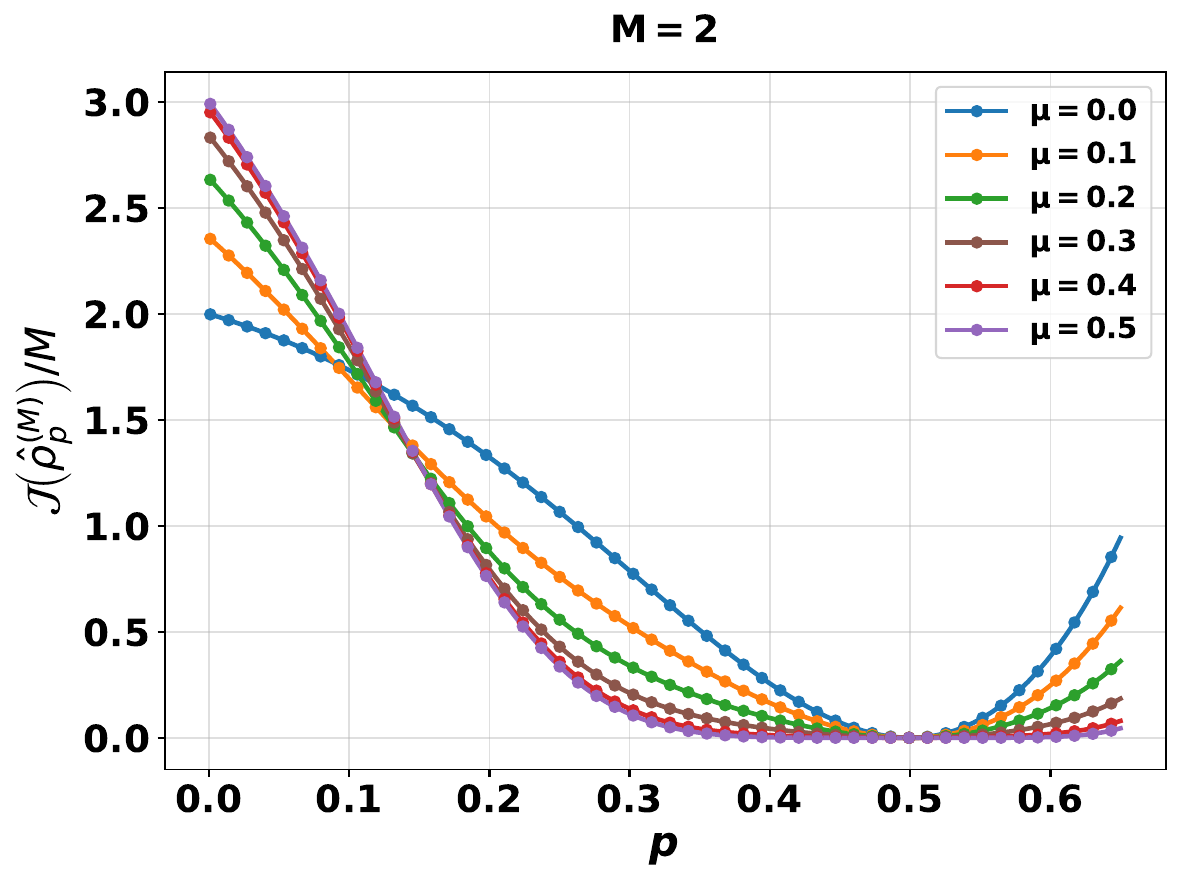}
    \caption{}
    \label{fig:1b}
\end{subfigure}
\hfill
\begin{subfigure}[b]{0.32\textwidth}
    \centering
    \includegraphics[width=\linewidth]{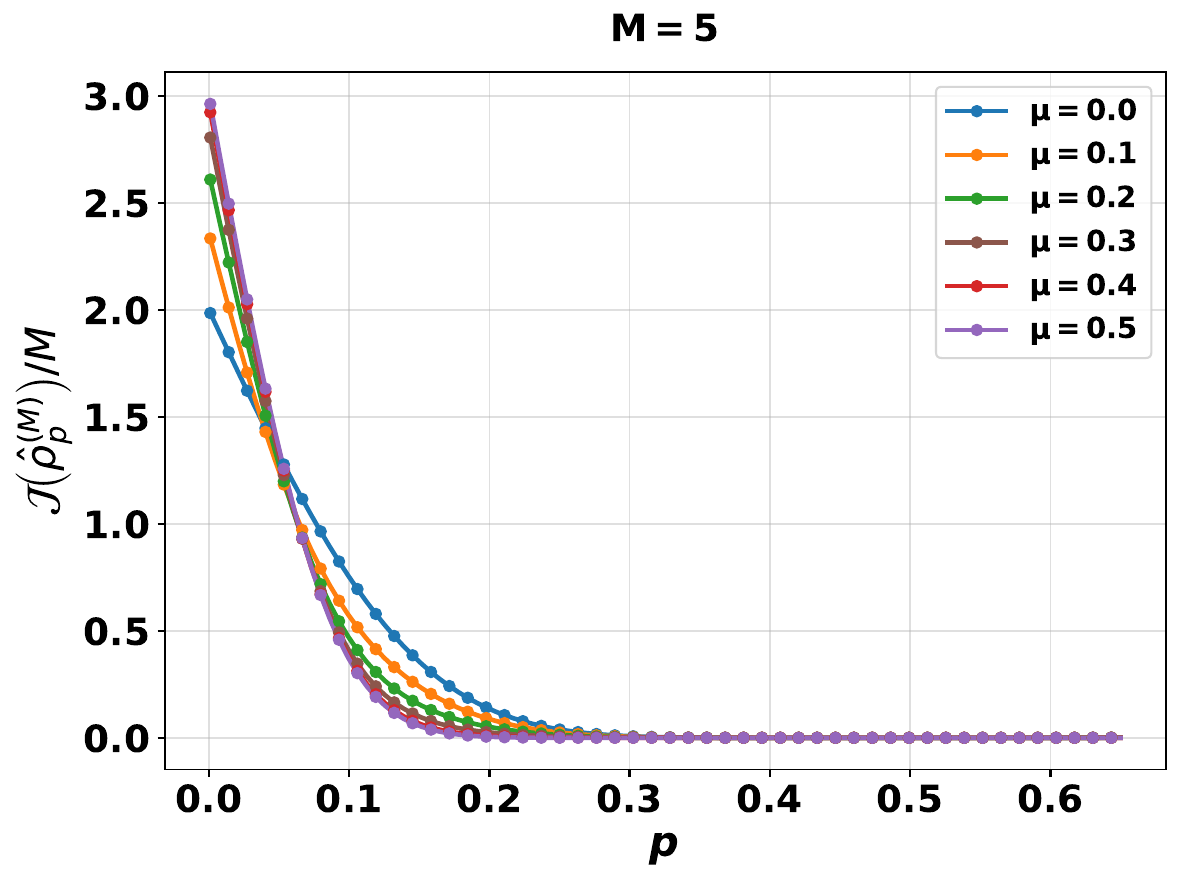}
    \caption{}
    \label{fig:1c}
\end{subfigure}
\caption{
RQFI under repeated applications of the two-qubit depolarizing channel for (a) $M=1$, (b) $M=2$, and (c) $M=5$, with $\mu$ ranging from $\mu=0$ (pure separable state) to $\mu=1/2$ (maximally entangled state). The plotted values of the RQFI are divided by $M$. For $M=1$, the RQFI remains strongly dependent on the initial state and shows a non-monotonic response for $\mu \ge 0.1$. For $M \ge 2$, the dependence 
of the RQFI on $\mu$ is significantly reduced, and the RQFI is vanishing over an increasing range of $p$. Already for $M=5$, the curves rapidly collapse toward zero and the differences between the curves corresponding to different $\mu$ become progressively smaller. As $M$ increases, the RQFI rapidly becomes insensitive to the initial two-qubit state, as the probes' state tends towards the completely mixed state. In the three panels, the numerical results are shown by markers connected with dashed lines, whereas the analytical predictions are represented by solid lines.
}
\label{fig:rqfi_three_panels}
\end{figure*}

In Fig.~\ref{fig:rqfi_three_panels} we plot the RQFI divided by $M$, as a function of $p$ for different values of $\mu$ and $M=1,2,5$ channel applications.
For $M=1$, panel (a), the RQFI depends strongly on $\mu$ and shows a non-monotonic response for $\mu\geq0.1$: it decreases in the interval $p \approx[0, 0.4{\rm -}0.5]$ until reaching the minimum; then, it increases again for larger values of $p$. The minimum of the RQFI is deepest for $\mu=0.5$ (maximally entangled state), where it approaches zero. The only exception is when $\mu=0$ (separable state), in which case the RQFI increases monotonically with $p$ over the entire range shown. After a second application of the channel ($M=2$, panel (b)), the dependence of the RQFI on $\mu$ is appreciably reduced: all curves decay rapidly towards zero for any value of $\mu$ considered. This decay is steeper in $p$ as $M$ increases, as shown in panel (c) corresponding to $M=5$. In such a case, the RQFI is practically zero for $p\geq0.2 {\rm -} 0.3$ irrespective of $\mu$. 

The behaviors plotted in the panels of Fig.~\ref{fig:rqfi_three_panels} follow from Eq.~\eqref{eq:Depol Chan}, in which the contribution of the initial state on $\Lambda_p^{M}(\hat\rho)$ is weighted by the factor $\xi_2(p)^{2M}$. This factor decreases with $p$ and decays exponentially with $M$, implying that increasing the number of channel applications progressively suppresses the dependence on the initial state, and drives the two-qubit probe towards the maximally mixed state at a rate that increases with $p, M$. Consequently, the metrological relevance of quantum correlations in the initial state encoded by $\mu$ is confined to a narrow range of small $p$ and $M$. Beyond this range, repeated depolarization erases the correlations that make the initial state informative about $p$.

\subsection{Optimal conditions for maximum QFI}

In Fig.~\ref{fig:rqfi_three_panels}, one can also observe that there exists a value of $p$ at which the behavior of the QFI as a function of $\mu$ changes radically: as $p$ increases, the optimal initial state of the two-qubit probe shifts from a maximally entangled state to a separable state.

\begin{figure}[t]
    \centering
    \includegraphics[width=\columnwidth]{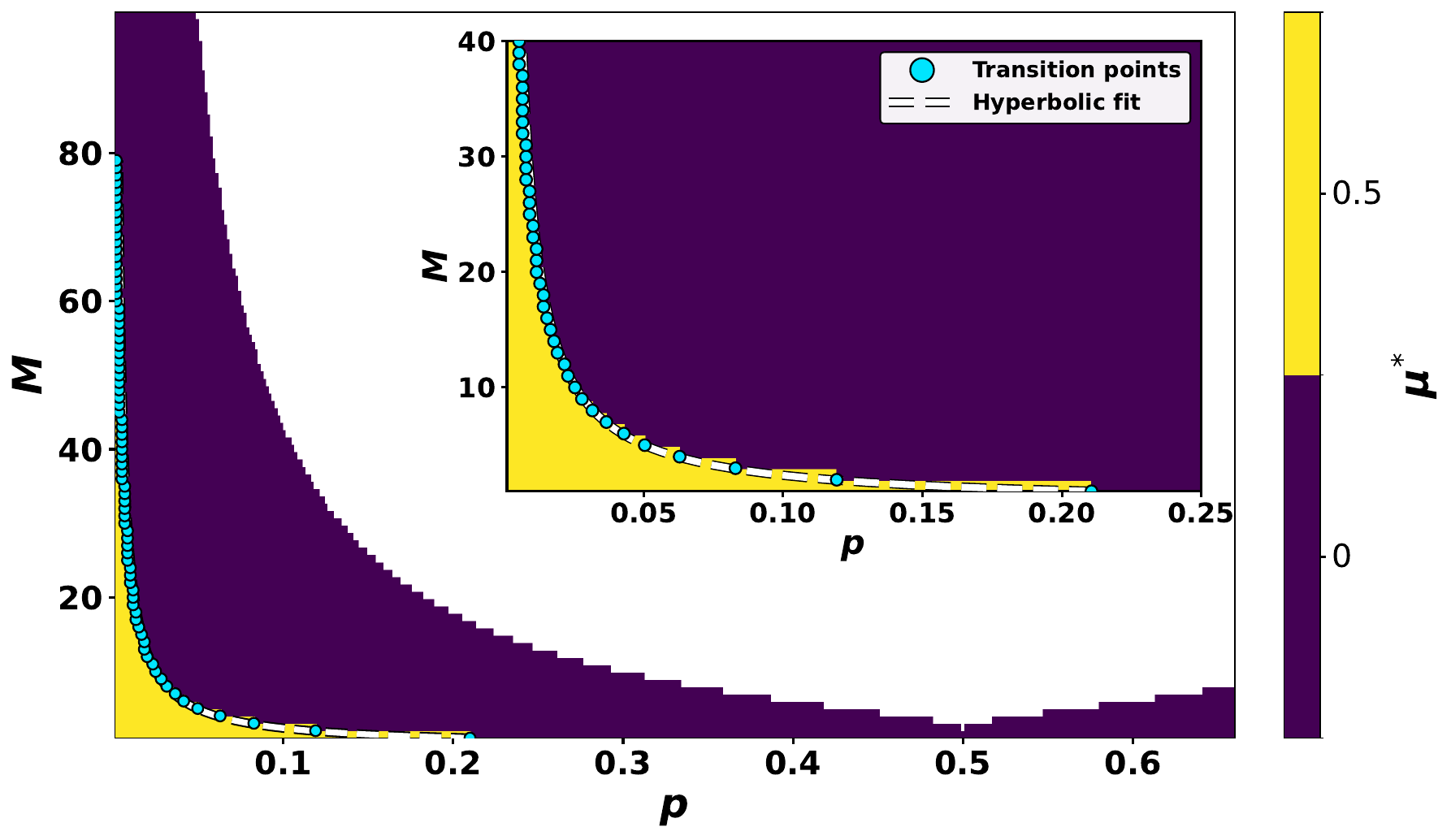}
    \caption{
    Optimal initial state parameter $\mu^*$, maximizing the QFI $J( \hat\rho_{p}^{(M)})$, under repeated channel applications with a QFI cut-off set to $10^{-4}$. The inset provides a higher-resolution view of the (low-$p$, low-$M$) regime. Two optimal regimes are determined: the separable state ($\mu^*=0$) and the maximally entangled state ($\mu^*=0.5$). In the weak-noise regime and small $M$, the QFI is maximized with $\mu^*=0.5$. As either $p$ or $M$ increases, the initial separable state becomes optimal, with the transition occurring at progressively smaller values of $p$ as $M$ increases. This behavior shows that, under repeated applications of the channel, the metrological advantage of maximal entanglement is confined to a progressively narrower area, delimitated by a  hyperbolic scaling curve. The latter is fitted through the function $M(p)= B/p-C$ with $B=0.263$ and $C=0.244$. 
    }
    \label{fig: Heatmap for Best mu}
\end{figure}

This effect is well-represented in the heatmap of Fig.~\ref{fig: Heatmap for Best mu}, where it is shown the value of $\mu$, $\mu^*$, that maximizes the QFI for the estimation of the depolarizing strength $p$, as a function of $p$ and $M$. The optimal $\mu$ happens to be exclusively one of the boundary values, $\mu^*=0$ and $\mu^*=0.5$, rather than one of the intermediate values. This indicates that the most informative two-qubit probe is either fully separable or maximally entangled depending on the operating point $(p, M)$. For a single application of the depolarizing channel, a similar transition was also observed in~\cite{Fujiwara_2001}, where it is found that, for the depolarizing strength estimation using a two-qubit probe, the initial state yielding maximum QFI is either a maximally entangled one, or the product of two pure states, depending on the depolarization parameter.  

For most points in the parameter space $(p, M)$, we find that the optimality is achieved by preparing the probes in a separable state (color purple in Fig.~\ref{fig: Heatmap for Best mu}), whereas the optimality is ensured by the maximally entangled preparation in the area of points highlighted in yellow. The latter is delimitated by a function with a hyperbolic scaling, whereby at large $M$ the preparation in an entangled state retains a comparative advantage for the probes' sensitivity in the weak-noise regime. This hyperbolic scaling marks a transition that reflects the competition between the enhanced parameter encoding achieved by repeated channel applications and the concurrent loss of quantum coherence and correlations due to depolarizing channels. The boundary marking the transition is fitted with the hyperbolic function $M(p)=B/p-C$ with $B=0.263$ and $C=0.244$. 

Notice that in Fig.~\ref{fig: Heatmap for Best mu} a region of points $(p, M)$ in the parameter space has not been colored. This comes from the fact that, as $p$ increases, the QFI decreases monotonically up to being vanishing, irrespective of $M$. In such a case, the value of the parameter $\mu$ becomes irrelevant and no a single optimal probe state can be determined. In other terms, the QFI vanishes for every value of $\mu$ and the choice of $\mu^*$ is degenerate. Therefore, in Fig.~\ref{fig: Heatmap for Best mu}, to ensure that the QFI remains meaningful within the parameter space, a cut-off value of $J_{\mathrm{cut-off}}=10^{-4}$ has been considered.

\begin{figure}[t]
    \centering
    \includegraphics[width=\columnwidth]{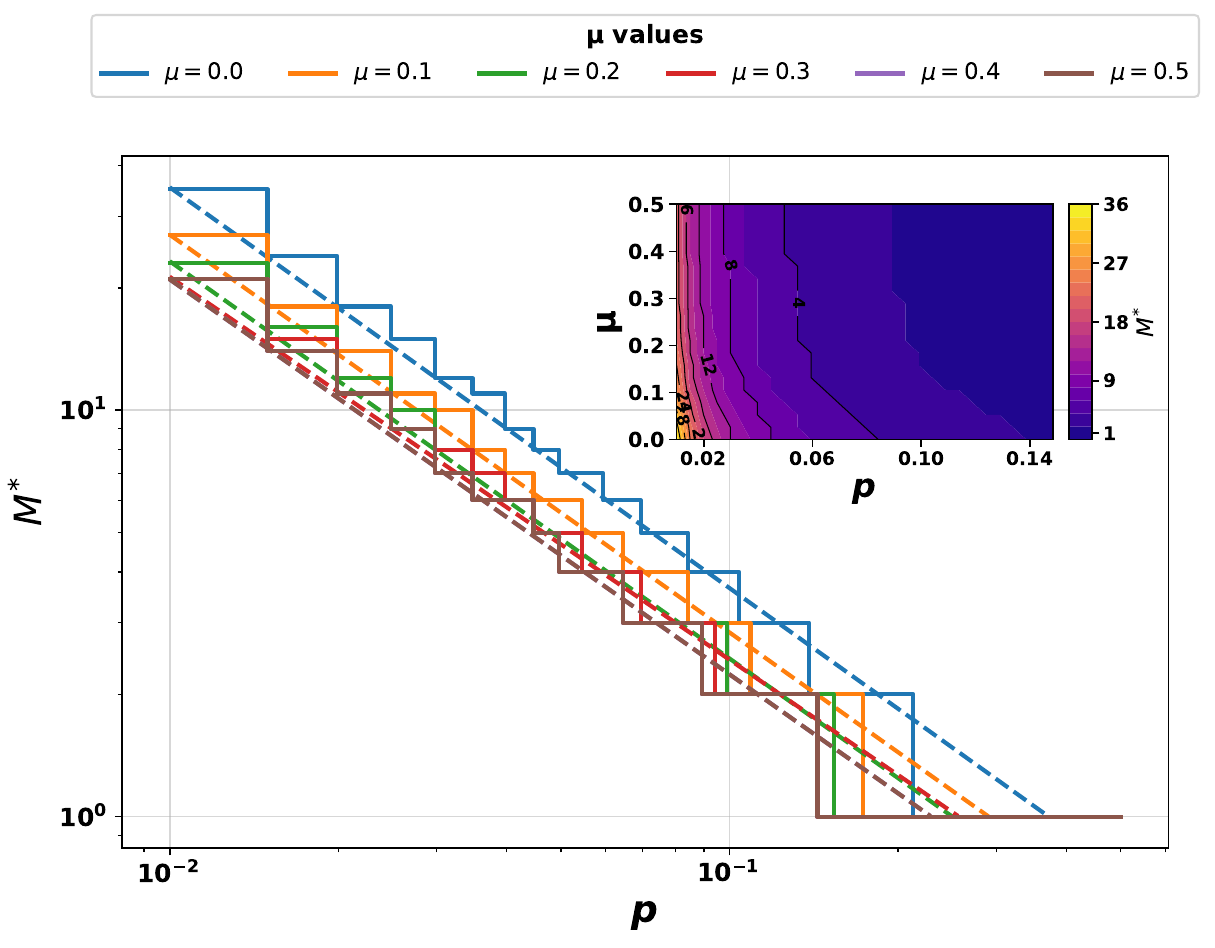} 
    \caption{
    Optimal number of repeated depolarizing channel applications, $M^*$, that allows for the maximization of the QFI $J( \hat\rho_{p}^{(M)})$. In the weak-noise regime, repeated applications of the channel enhance the sensitivity to $p$. With increasing depolarizing strength, repeated channel applications progressively degrade quantum coherence/correlations of the probes' state. Thus, the optimal strategy rapidly reduces to fewer repetitions, reaching eventually $M=1$. The log-log scale reveals the power-law nature of the decay $\sim 1/p$. The dependence on the initial state parameter $\mu$ is most visible at small $p$, while for stronger depolarizing strength all curves collapse to the single-application optimum $M^*=1$. \textit{Inset:} Contour map view of $M^*$ in the $(p, \mu)$ parameter space, which emphasizes the region of $p$ where $M^*$ changes the most. The larger values of $M^*$ are mainly at small $p$, while increasing $p$ rapidly leads to a small $M^*$ across different values of $\mu$.}
    \label{fig: Best M_vs_p heatmap}
\end{figure}

We now show in Fig.~\ref{fig: Best M_vs_p heatmap} the optimal number of depolarizing channel applications, $M^*$, that maximizes the QFI by varying $p$ and $\mu$. As in Fig.~\ref{fig:Optimal_M_function_of_p}, a staircase trend for $M^*$ is determined using the log-log scale, which arises as $M$ is an integer-valued variable. In fact, as $p$ changes continuously, the optimum $M$ can only jump between discrete values rather than vary smoothly. 

In the weak-noise regime, a decay $\sim 1/p$ is determined: $M^*$ takes large values, exceeding $M=30$ as $p\to 0$, and drops rapidly as $p$ increases. This behavior, as above, reflects the competition between two opposing effects: each additional application enhances the response of the probes' state to small changes in $p$, while suppresses quantum coherence and correlations that support the probes' sensitivity. For very small $p$, information accumulates faster than correlations decay, which means that a large number of channel applications are required before this trade-off is reached. Beyond the weak-noise regime, a single channel application can be already sufficient for optimal estimation.

The dependence of $M^*$ on $\mu$ is most pronounced in the weak-noise regime. Among the values considered, the separable configuration $\mu=0$ allows for the largest $M^*$, whereas increasing $\mu$ systematically lowers it, indicating that high-correlated initial states are more affected by depolarization just after few applications. As $p$ increases, this dependence on $\mu$ weakens rapidly, and the curves corresponding to different $\mu$ collapse onto a common $M^*$ value. Therefore, for large value of $p$, the optimal number of channel applications is essentially independent of how the probe was initially prepared.
The heatmap in the inset of Fig.~\ref{fig: Best M_vs_p heatmap} confirms this interpretation: large values of $M^*$ are confined to a small region near $p\to 0$, while over most of the parameter space $(p, \mu)$ it is worth consider only a few applications of the channel and eventually a single application.

\subsection{RQFI slope analysis in the $p \to 0$ limit}
\label{sec:slope}

In this section, we aim to study how the decay of the RQFI, in its early time-behavior, depends on the amount of correlations in the initial state of the probes, under repeated applications of the channels. To do this, we analyze the derivative of the RQFI with respect to $p$ evaluated at $p=0$, $\partial_p \mathcal{J}( \hat\rho_p^{(M)} )|_{p=0}$, for $\mu=0,0.1,\ldots,0.5$ and $M=1,\ldots,6$. This quantity provides a measure of the system's initial sensitivity to changes in the depolarizing parameter and allows us to compare how initial states with different correlations respond to noise.

\begin{figure}[t]
    \centering
    \includegraphics[width=0.95\columnwidth]{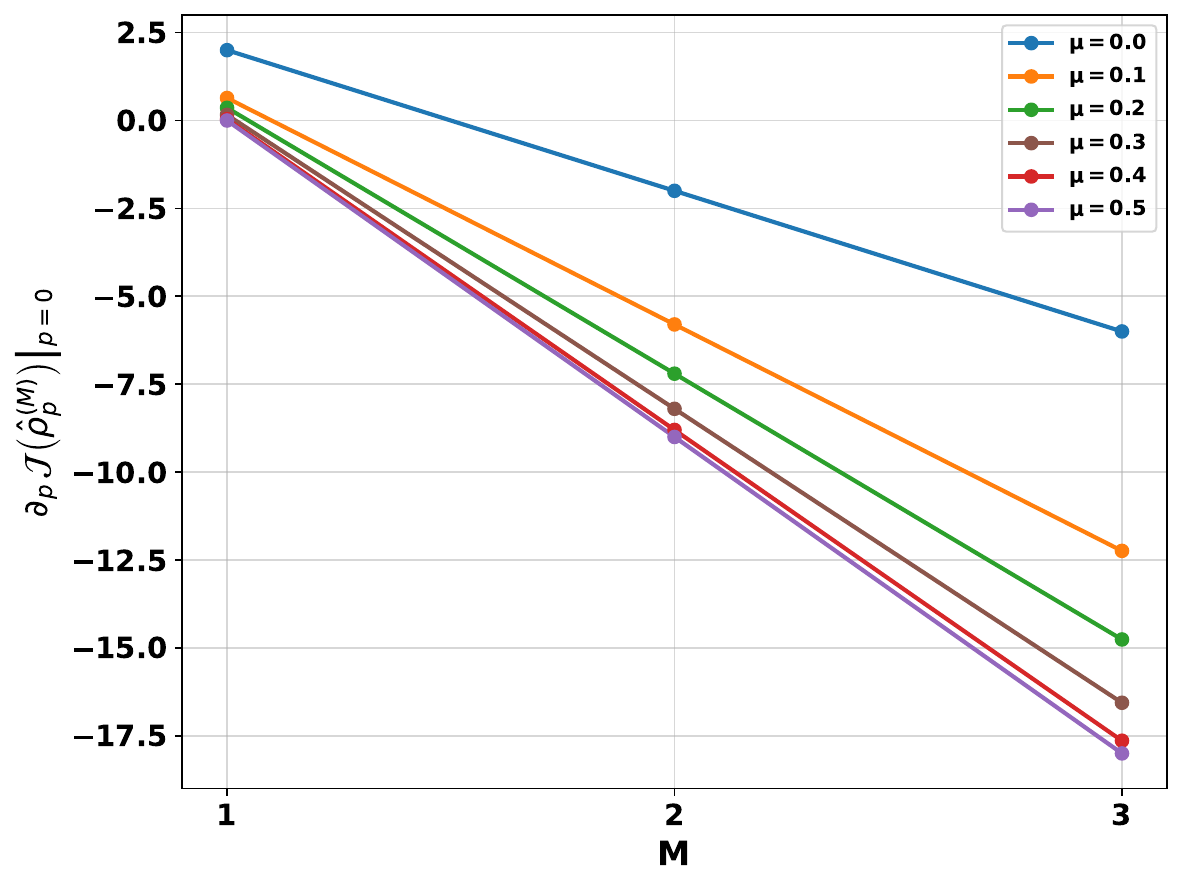} 
    \caption{
    Plot of $\partial_p \mathcal{J}( \hat\rho_p^{(M)} )|_{p=0}$, representing the slope of RQFI at $p=0$, as a function of the number $M$ of depolarizing channel applications for $\mu=0,0.1,\ldots,0.5$. The increasingly negative slopes of RQFI as $M$ grows means that repeated channel applications accelerate the loss of metrological sensitivity. Moreover, the steepest decay observed for increasing $\mu$ indicates that an initially more correlated two-qubit probe is more sensitive to noise degradation. This reveals a trade-off between correlation-enhanced precision and robustness under depolarization. The numerical results are shown by markers connected with dashed lines, whereas the analytical predictions (see Appendix \ref{sec:slope_RQFI}) are represented by solid lines.
    }
    \label{fig: Slope vs M}
\end{figure}

$\partial_p \mathcal{J}( \hat\rho_p^{(M)} )|_{p=0}$, corresponding to the slope of the RQFI for $p\to 0$, becomes increasingly negative as the number of channel applications $M$ increases, as shown in Fig.~\ref{fig: Slope vs M} where the initial state of Eq.~\eqref{eq: initial state} is considered. For $M=1$, the slope remains positive across the values of $\mu$ considered, showing that with a single application of the channel the sensitivity of the probes' state about $p$ is preserved. Then, for $M=2$ and $M=3$, the slope is negative for all $\mu$, indicating that the RQFI starts decreasing rapidly as soon as depolarizing noise affects the probe. In addition, larger the correlation magnitude $\mu$ and steepest is the slope of the RQFI at $p=0$. This means that the initial state with $\mu=0$ is the most stable as $p$ increases. In contrast, states with $\mu>0$, especially when approaching $\mu=0.5$, lose sensitivity faster when $p$ moves away zero. Therefore, although quantum correlations in the initial state enhances the sensitivity of the probe about $p$, they also make the probe more fragile under repeated applications of the noise channel. 
  
\begin{figure}[t]
    \centering
    \includegraphics[width=0.95\columnwidth]{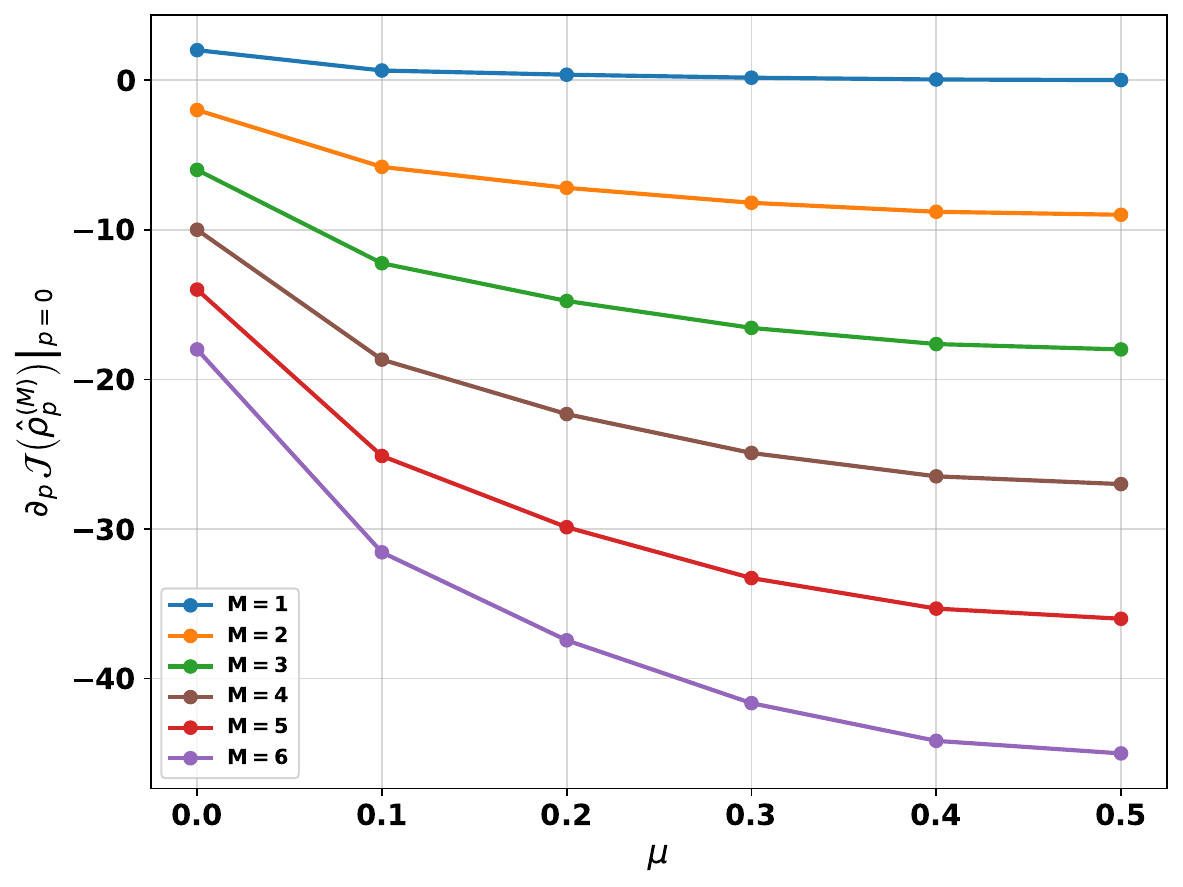} 
    \caption{
    $\partial_p \mathcal{J}( \hat\rho_p^{(M)} )|_{p=0}$ as a function of $\mu$, for $M=1,\ldots,6$. As $\mu$ grows, the slope of the RQFI at $p=0$ becomes progressively more negative, and this trend is more pronounced with an increasing number $M$ of channel applications. The numerical results are shown by markers connected with dashed lines, whereas the analytical predictions (see Appendix \ref{sec:slope_RQFI}) are represented by solid lines.
    }
    \label{fig: Slope vs mu}
\end{figure}

Fig.~\ref{fig: Slope vs mu} highlights the behavior of $\partial_p \mathcal{J}( \hat\rho_p^{(M)} )|_{p=0}$ as a function of $\mu$, for $M=1,\ldots,6$. For every $M$, $\partial_p \mathcal{J}( \hat\rho_p^{(M)} )|_{p=0}$ decreases monotonically as $\mu$ increases from $0$ to $0.5$. In addition, larger $M$, and steeper is $\partial_p \mathcal{J}( \hat\rho_p^{(M)} )|_{p=0}$ with respect to $\mu$. Therefore, both repeated channel applications and the amount of correlations in the initial state of the probe accelerate the loss of parameter sensitivity. Such decay behaviors are similar with those identified in Fig.~\ref{fig: Slope vs M}, albeit following a slight different trend. In conclusion, an initially correlated probe provides an enhanced metrological response, but their sensitivity to the parameter of interest is degraded more rapidly when the channel is applied multiple times.  

\section{Conclusions}
\label{sec:conclusion}

In this work, we have systematically studied the sequential strategy for estimating the strength of a depolarizing channel, using single and two-qubit probes. In particular, using the QFI and RQFI to quantify the estimation precision, we have investigated to which extent the number $M$ of depolarizing channel applications can be regarded as an additional metrological resource. Our studies are relevant, since the accurate characterization of noise is crucial for maximizing the reliability of quantum technologies. In particular, depolarizing noise needs to be precisely measured in noisy quantum processors due to its impact on the accuracy of quantum computation. A recent research~\cite{Pau_Escofet} highlights the need of simulating depolarizing noise for circuit fidelity estimation, hardware benchmarking, and designing larger quantum systems.

We have demonstrated that, in the weak-noise regime, repeated channel applications amplify the QFI, and this amplification linearly increases with $M$ in the limit $p \to 0$. Pure input states provide the largest QFI. The behavior changes as the depolarizing strength increases, whereby the RQFI reaches its maximum at an optimal value of $M$, $M^*$, indicating in such a context that an arbitrarily large value of $M$ is no longer beneficial. Furthermore, the RQFI decays faster with $p$ as $M$ increases, highlighting that repetition is mostly useful in the weak-noise regime. 
Furthermore, we have compared the single-qubit sequential protocol with the extended optimal strategies, employing a two-qubit probe. This study shows that in certain weak-noise scenarios, repetition can improve the sensitivity over extended single-round approaches. The advantage, however, is not uniform and becomes less effective as the depolarizing strength increases. This indicates that the number of channel applications should be optimized rather than just increased.

For two-qubit probes, the estimation precision depends on the correlations in the initial state, which we have chosen equal to $|\psi\rangle = \sqrt{\mu}|00\rangle+\sqrt{1-\mu}|11\rangle$, with $\mu$ tuning the amount of correlations in the state. We have thus investigated the values of $M$ and $\mu$, $M^*$ and $\mu^*$, that maximize the RQFI. For small values of $p$, the maximal RQFI is reached for $\mu^*=1/2$ corresponding to a maximally entangled state, irrespective of the value of $M$. Instead, as the depolarizing strength $p$ increases, the optimal number of repetitions $M^*$ progressively decreases towards $M^*=1$ and the dependence on $\mu$ loses relevance, since $\mu^*=0$ (associated to an initial separable state) yields the maximum QFI for $p$ sufficiently large. However, albeit the preparation of the probes in an initially correlated state make them more sensitive to $p$ in the weak-noise regime, such sensitivity degraded more rapidly, especially when the depolarizing channel is applied multiple times.  

Our results suggest possible approaches for experimental implementation. Photonic polarization qubits offer a viable platform, as depolarizing noise and entangled two-photon probes have already been studied in quantum metrology experiments~\cite{Stephen_Walborn, ShahamPRA2011, JeongPRA2013, Wang2018, PhysRevA.103.012602}. Specifically, in both \cite{Wang2018} and \cite{PhysRevA.103.012602}, it has been investigated phase estimation in the presence of depolarizing noise using polarization-entangled photons. The latter study shows that controlled depolarizing noise could be implemented experimentally in a photonic metrological system. Hence, our results about weak-noise QFI enhancement with $M > 1$ could be tested experimentally by repeatedly applying a depolarizing channel to a single or two-qubit probe.

In addition to an experimental implementation of our findings using the sequential strategy of metrology, we have identified the following outlooks. (i) First, it would be interesting to expand the current study to include more than two probes (thus, $N>2$). This would allow us to figure out how the optimal number of channel repetitions changes with the probe size, as well as if the behaviors observed with few probes remain valid in larger systems. (ii) A further direction is to further explore the relationship between the RQFI and multipartite entanglement, by quantifying the capability of a sequence of $M$ depolarizing channels to detect entanglement~\cite{SuzukiPRA2016}. (iii) Another natural extension is to analyze quantum metrology performance of estimation the depolarizing strength in noisy quantum circuits where unitary operators are deteriorated by depolarizing noise. (iv) The sequential metrological protocol might also be applied to quantum sensor networks and distributed sensing~\cite{MalitestaPRA2023,PezzePRL2025}, studying how repeated local channel applications and correlations between spatially distributed probes affect the estimated precision.

\section*{Data availability statement}

The data plotted in the Figs.~\ref{fig:Generic_d_Generic_M}--\ref{fig: Slope vs mu} are available at~\cite{Yasir2026}.

\section*{Acknowledgments}

The authors acknowledge very useful discussions with Gabriel Aguilar, Beatrice Donelli and Marco Malitesta.
This work has been financial supported by the PNRR MUR project PE0000023 NQSTI funded by the European Union---Next Generation EU, and the European project ``High Performance Computer and Quantum Simulator hybrid (HPCQS)'' with grant agreement No. 101018180 funded by the European Union's Horizon 2020 research and innovation programme.

\appendix

\section{Single quantum probe}
\label{sec:SQP}

In this appendix, we present the derivation of the QFI for a single pure quantum probe, with generic dimension $d$, subjected to a depolarizing channel with strength $p$. This derivation provides the analytical framework to characterize the sensitivity of the probe's state to the noise parameter $p$.

\subsection{Derivation of QFI for a qudit probe}

\subsubsection{Single channel application}

Using the notation $\xi_d(p) := (1-pd)$, let us consider the expression of a depolarizing channel applied to a pure qudit probe with ${\rm Tr}[\hat\rho]={\rm Tr}[\hat\rho^2]=1$:
\begin{align}
    \hat\rho_{p}=\Lambda_p(\hat\rho)=\xi_d(p)\hat\rho + p\hat{\id}_d.
    \label{eq:DPC}
\end{align}
Defining the projector $\hat{Q}=\hat\id_d-\hat\rho$, such that $\hat\rho\hat Q=\hat Q\hat\rho=0$, $\hat\rho_{p}$ can be equivalently expressed as
\begin{equation}
    \hat\rho_p = [1-(d-1)p]\hat\rho + p\hat{Q}.
    \label{eq:a3}
\end{equation}
The goal now is to solve the Lyapunov equation, Eq.~\eqref{eq:lyapunov} in the main text. Let us take the derivative of $\hat\rho_p$ with respect to $p$:
\begin{equation}
    \partial_p\hat\rho_p = -(d-1)\hat\rho + \hat{Q} = \hat{\id}_d -\hat{\rho} d.
\end{equation}
Then, we parametrize also the SLD as a function of $\hat{\rho},\hat{Q}$:
\begin{equation}
    \hat{L}_p = a(p)\hat{\rho} + b(p)\hat{Q},
    \label{eq:a5}
\end{equation}
whose coefficients $a(p)$ and $b(p)$ have to be determined. For such a purpose, we compute the anti-commutator $\{ \hat{L}_p, \hat{\rho}_p \}$. Using the relation $\hat{\rho}\hat{Q} = \hat{Q}\hat{\rho} = 0$, it follows that
\begin{equation}
    \hat{\rho}_p \hat{L}_p = \hat{L}_p \hat{\rho}_p = a(p)[1-(d-1)p]\hat\rho+b(p)p \hat Q.
\end{equation}
Therefore, by substituting into the Lyapunov equation, we obtain the expressions for $a(p)$ and $b(p)$ by equating the coefficients in front of each operator:
\begin{equation}
    a(p) = \frac{-(d-1)}{1-(d-1)p},
    \qquad
    b(p) = \frac{1}{p}.
\end{equation}
Hence, the SLD reads as
\begin{equation}
    \hat{L}_p = \frac{-(d-1)}{1-(d-1)p}\hat{\rho} + \frac{1}{p}\hat{Q}.
    \label{eq:SLD}
\end{equation}
As a result, the QFI can be evaluated through Eq.~\eqref{eq: QFI_rhop} as
\begin{equation}
J(\hat{\rho}_{p}) = \Tr\left[\hat{L}_p^2 \hat{\rho}_p\right] = \frac{(d-1)^{2}}{1-(d-1)p}+\frac{(d-1)}{p}.
\label{eq:unexpanded_qfi}
\end{equation}
In the weak-noise regime, we can expand the first term of Eq.~\eqref{eq:unexpanded_qfi} in powers of $p$, namely
\begin{equation}
    \frac{(d-1)}{1-(d-1)p} = d-1 +(d-1)^{2}p+2(d-1)^{3}p^{2}+O(p^{3}).
\end{equation}
In this way, we get
\begin{align}
    J(\hat\rho_{p})=\frac{(d-1)}{p}+(d-1)^2+(d-1)^{3}p+O(p^{2}),
\end{align}
which represent the QFI of the final state of a single qubit with respect to the parameter $p$, encoded through a single application of the depolarizing channel, in the weak-noise regime.
\subsubsection{Multiple applications of the channel}

We now extend the analysis to a generic number $M$ of channel applications, after which the qudit probe state is
\begin{align}    
\hat{\rho}^{(M)}_{p} = \xi_d(p)^{M}\hat{\rho} + \left( 1-\xi_d(p)^{M} \right)\frac{\hat{\id}_d}{d},
\label{eq:rho_p_M}
\end{align}
or equivalently
\begin{equation}
    \hat{\rho}^{(M)}_{p} = \frac{1+(d-1)\xi_d(p)^{M}}{d}\hat{\rho} + \frac{(1-\xi_d(p)^{M})}{d}\hat{Q}.
    \label{eq:rho_p_M_equiv}
\end{equation}
To solve the Lyapunov equation, we take the derivative of $\hat{\rho}_{p}^{(M)}$ with respect to the depolarizing strength $p$:
\begin{equation}
\partial_{p}\hat\rho^{(M)}_{p}
    =-(d-1)M\xi_d(p)^{M-1}\hat\rho+M\xi_d(p)^{M-1}\hat Q.
\end{equation}
Then, as before, we parametrize the SLD as a function of $\hat{\rho}, \hat{Q}$, so that 
\begin{equation}
    \hat{L}^{(M)}_{p}=a_{M}(p)\hat{\rho} + b_{M}(p)\hat{Q}.
    \label{eq:SLD for M}
\end{equation}
From solving the Lyapunov equation, we determine that $a_{M}$ and $b_{M}$ are respectively:
\begin{equation}
\begin{split}
    &a_{M}(p) = -\frac{d(d-1)M\xi_d(p)^{M-1}}{1+(d-1)\xi_d(p)^M}, 
    \\
    &b_{M}(p) = \frac{M\xi_d(p)^{M-1}d}{1-\xi_d(p)^M}.
    \label{eq: AMBM}
\end{split}
\end{equation}
As a result, the QFI is
\begin{align}
J\left(\hat\rho_{p}^{(M)}\right)=\frac{d^2(d-1)M^{2}\xi_d(p)^{2(M-1)}}{(1+(d-1)\xi_d(p)^{M})(1-\xi_d(p)^{M})}.
\label{eq:qfi_sq_mrep}
\end{align}

\section{Two-qubit probe}
\label{sec:two_qubit_depol_derivation}

In this section, we present a detailed derivation of the QFI in estimating the depolarization strength of a depolarizing channel applied repeatedly ($M$ times) to a two-qubit system. In particular, the goal is to provide a complete analytical formulation for the numerical results discussed in section ~\ref{sec:depol_derivation}.

First, the evolution of the initial pure two-qubit state $\hat{\rho}$ under a single application ($M=1$) of the depolarizing channel to each qubit, with depolarizing parameter $p$, is described by
\begin{align}
    \hat{\rho}_p =\Lambda_p(\hat\rho) = \xi_2(p)^{2}\hat{\rho}+p\xi_2(p)S(\hat\rho)+p^{2}\hat\id_{4},
    \label{eq: FAM1}
\end{align}
where 
\begin{equation}
    S(\hat\rho) :=\hat\id_2\otimes\Tr_{Q_1}[\hat\rho] + \Tr_{Q_2}[\hat\rho]\otimes\hat\id_2.
    \label{eq:S_def}
\end{equation}
We now generalize Eq.~\eqref{eq: FAM1} to an arbitrary number $M$ of channel applications. First, from the definition of $S(\hat\rho)$ in Eq.~\eqref{eq:S_def}, we obtain the relations
\begin{equation}
S(S(\hat\rho)) = 2S(\hat\rho) + 2\hat\id_4, 
\quad
S(\hat\id_4)=4\hat\id_{4},
\label{eq:SSrho}
\end{equation}
by imposing the equality of the two single-qubit reduced states, namely $\Tr_{Q_1} [\hat{\rho}] = \Tr_{Q_2} [\hat{\rho}]$, which is guaranteed by the invariance of $\hat\rho$ under qubit exchange.
As $\mathrm{Tr}[S(\hat\rho)]=\mathrm{Tr}[\hat\id_4]=4$, the action of the channel on $S(\hat\rho)$ is
\begin{equation}
\Lambda_p(S(\hat\rho))
=
\xi_2(p)S(\hat\rho)+2p\hat\id_4,
\label{eq:LambdapSrho}
\end{equation}
while leaving the identity operator unchanged:
\begin{equation}
    \Lambda_p(\hat\id_4)=\hat\id_4.
    \label{eq:LambdapId}
\end{equation}
Hence, under repeated action of the channel, the evolved state remains in the linear span of the three operators $\hat\rho$, $S(\hat\rho)$ and $\hat\id_4$. 
Thus, we parametrize the final state as
\begin{equation}
\hat{\rho}_p^{(M)} 
=
A_M\hat\rho + B_M S(\hat\rho)+C_M \hat\id_4 ,
\label{eq:Mapplications}
\end{equation}
where $A_M$, $B_M$, and $C_M$ are scalar coefficients depending on $M$ and $p$ to be determined. Using mathematical induction, we obtain the following expressions for these coefficients:
\begin{equation}
\begin{aligned}
    A_M &= \xi_2(p)^{2M}, \\
    B_M &=\frac{\xi_2(p)^M\left[1-\xi_2(p)^{M}\right]}{2}, \\
C_M &=\left(\frac{\xi_2(p)^M -1}{2} \right)^2.
\end{aligned}
\label{eq:coefficients_rho_twoqubits_Mapplications}
\end{equation}
Indeed, for $M=1$, substituting these coefficients into Eq.~\eqref{eq:Mapplications} reproduces Eq.~\eqref{eq: FAM1}. Then, assuming that these relations hold for a given $M$, using Eqs.~\eqref{eq: FAM1}, \eqref{eq:LambdapSrho} and \eqref{eq:LambdapId} we can evaluate
\begin{align}
    \hat{\rho}_p^{(M+1)} 
= & \xi_2(p)^{2M} \Lambda_p (\hat{\rho}) + \frac{\xi_2(p)^M\left[1-\xi_2(p)^{M}\right]}{2} \Lambda_p (S(\hat{\rho})) \notag\\
&+\left(\frac{\xi_2(p)^M -1}{2} \right)^2 \Lambda_p (\id_4) \notag \\
 = & A_{M+1}\hat\rho + B_{M+1} S(\hat\rho)+C_{M+1} \hat\id_4.
\end{align}
This completes the proof. Substituting Eq.~\eqref{eq:coefficients_rho_twoqubits_Mapplications} into Eq.~\eqref{eq:Mapplications}, we obtain the expression for the state after $M$ applications of the depolarizing channel:
\begin{align}
    \hat{\rho}_p^{(M)} = & \: \xi_2(p)^{2M}\hat\rho + \frac{\xi_2(p)^M\left[1-\xi_2(p)^{M}\right]}{2} S(\hat{\rho})\notag\\
    &+\frac{1-2\xi_2(p)^{M}+\xi_2(p)^{2M}}{4}\hat\id_{4},
    \label{eq: M depolarizing channel}
\end{align}
with $S(\hat{\rho})$ defined as in Eq.~\eqref{eq:S_def}.
This closed-form expression allows us to systematically investigate the decay of sensitivity under repeated applications of the depolarizing channel and to quantify the trade-off between enhanced precision and noise-induced fragility.

\subsection{QFI in the weak-noise regime}
\label{sec:app_two_qubit_weak_noise}

In this subsection, we derive the QFI for a generic two-qubit pure input state subjected to M applications of the depolarizing channel, in the weak-noise regime. Specifically, we assume that the depolarizing strength $p$ is small, and retain only the leading order contributions in $p$. 
To this end, we expand the coefficients of Eq.~\eqref{eq: M depolarizing channel} in a Taylor series around $p=0$: 
\begin{equation}
    \xi_2(p)^M = 1-2pM+2M(M-1)p^2+O(p^3),
\end{equation}
and similarly
\begin{equation}
    \xi_2(p)^{2M} = 1-4pM+4M(2M-1)p^2+O(p^3).
\end{equation}
Hence, the final state can can be written up to $O(p^3)$ as
\begin{equation}
\hat{\rho}_p^{(M)} = \hat\rho + p \hat\rho_1^{(M)} + p^2 \hat\rho_2^{(M)} + O(p^3),
\label{eq:Lambda_M_2ndorder}
\end{equation}
while the derivative of \eqref{eq:Lambda_M_2ndorder} reads
\begin{equation}
    \partial_p \hat{\rho}_p^{(M)} = \hat\rho_1^{(M)} + 2 p \hat\rho_2^{(M)} + O(p^2).
\label{eq:DerivativeM}
\end{equation}
We defined $\hat\rho_j^{(M)} := (\partial_p)^j \hat\rho_p^{(M)}|_{p=0}$, which explicitly gives
\begin{align}
    \hat\rho_1^{(M)} &=  M (S - 4\hat\rho) ,
    \label{eq:rho_1_M}
    \\
    \hat\rho_2^{(M)} &= 4M(2M-1)\hat\rho-M(3M-1)S(\hat\rho)+M^{2}\hat\id_{4}.\notag
\end{align}
The expansion of Eq.~\eqref{eq:Lambda_M_2ndorder} allows for the evaluation of the SLD in the weak-noise regime. In particular
\begin{align}
\hat{L}_p^{(M)} = \frac{\hat\id_4 - \hat\rho}{p} - \hat\rho_1^{(M)} + O(p).
\label{eq:SLD}
\end{align}
This can be readily checked by inserting Eqs.~\eqref{eq:Lambda_M_2ndorder}, \eqref{eq:DerivativeM} and \eqref{eq:SLD} in the Lyapunov equation~\cite{Hota2005,Boixo_2008}.

As $\Tr[ \hat{\rho}_p^{(M)} \hat L_p^{(M)}] = 0 $, the QFI can be equivalently defined as
\begin{equation}
J\left( \hat\rho_p^{(M)} \right) = \mathrm{Tr}\left[ \left(\partial_p \hat{\rho}_p^{(M)} \right) \, L_p^{(M)} \right].
\label{eq:QFIM}
\end{equation}
Thus, using Eqs.~\eqref{eq:DerivativeM} and \eqref{eq:SLD}, we obtain
\begin{equation}
J\left( \hat\rho_p^{(M)} \right) = \frac{1}{p} \mathrm{Tr}\left[ \hat\rho_1^{(M)} (\hat{\id}_4 - \hat\rho) \right] + O(1).
\label{eq:JM}
\end{equation}

The leading order of the QFI in the weak-noise regime depends on $\hat\rho_1^{(M)}$, whereas the dependence on $\hat\rho_2^{(M)}$ appears only at higher orders, which have been neglected.
From Eq.~\eqref{eq:rho_1_M}, it follows directly that $\hat\rho_1^{(M)} = M \hat\rho_1^{(1)}$. Consequently,
\begin{equation}
    J\left( \hat\rho_p^{(M)} \right) = M J\left( \hat\rho_p^{(1)} \right) + O(1),
\end{equation}
showing that, in the weak-noise regime, the leading order of the QFI scales linearly with the number of channel applications $M$.

Moreover, from Eq.~\eqref{eq:rho_1_M}, we can evaluate
\begin{equation}
\mathrm{Tr}\left[ \hat\rho_1^{(M)} (\hat{\id}_4 - \hat\rho) \right] = M \left( 4 - P_{Q_1} - P_{Q_2} \right),
\label{eq:BM}
\end{equation}
where $P_{Q_1} = P_{Q_2} = P$ are the purities of the (identical) single-qubit reduced states of $\hat{\rho}$.
Substituting~\eqref{eq:BM} into~\eqref{eq:JM}, we get 
\begin{equation*}
J\left( \hat\rho_p^{(M)} \right)  = \frac{M (4 - 2P)}{p} + O(1).
\end{equation*}
The RQFI in the $p \to 0$ limit is then
\begin{equation}
\lim_{p\to 0} \mathcal{J}\left( \hat\rho_p^{(M)} \right) = M(4-2P),
\end{equation}
which makes the linear scaling in $M$ explicit.
\newline

\subsection{QFI for the initial state \eqref{eq: initial state} and arbitrary $p$}
\label{app:qfi_for_specific_two_qubit_initial_state} 

In this section, we derive an analytical expression for the QFI of a two-qubit system initialized in the pure state of Eq.~\eqref{eq: initial state} subjected to $M$ repeated depolarizing channel applications. The corresponding initial density matrix in the computational basis can be written as
\begin{equation}
  \hat\rho
  =
  \begin{pmatrix}
    \mu & 0 & 0 & \sqrt{\mu(1-\mu)}   \\
    0   & 0 & 0 & 0   \\
    0   & 0 & 0 & 0   \\
    \sqrt{\mu(1-\mu)}   & 0 & 0 & 1-\mu
  \end{pmatrix}.
  \label{eq:rhoinit}
\end{equation}

Note that the only non-zero entries lie in the $\{\ket{00},\ket{11}\}$ subspace. The final state has already been derived in Eq.~\eqref{eq: M depolarizing channel}. We proceed with the parametrization as in Eq.~\eqref{eq:Mapplications} for convenience, with coefficients $A_M$, $B_M$ and $C_M$ defined as in Eq.~\eqref{eq:coefficients_rho_twoqubits_Mapplications}. For the initial state given in Eq.~\eqref{eq:rhoinit}, the corresponding operator $S(\hat\rho)$ takes the explicit form
$S(\hat\rho)= \text{diag} \: \left( 2\mu, \: 1, \: 1, \: 2(1-\mu)\right)$. Hence, by substituting Eq.~\eqref{eq:rhoinit} and the expression of $S(\hat\rho)$ into Eq.~\eqref{eq:Mapplications}, $\hat\rho_p^{(M)}$ becomes
\begin{widetext}
    \begin{equation}
\hat\rho_p^{(M)}
=
\begin{pmatrix}
\mu D_M + C_M & 0 & 0 & \sqrt{\mu(1-\mu)} A_M \\
0 & B_M+C_M & 0 & 0 \\
0 & 0 & B_M+C_M & 0 \\
\sqrt{\mu(1-\mu)} A_M & 0 & 0 & (1-\mu) D_M + C_M
\end{pmatrix},
\label{eq:rhoout}
\end{equation}
\end{widetext}
where we defined $D_M := A_M + 2 B_M = \xi_2(p)^M$.

We now compute the QFI analytically by means of its spectral decomposition as in Eq.~\eqref{eq:master}. Since the initial state $\hat\rho$ is independent of $p$, differentiating Eq.~\eqref{eq:Mapplications} with respect to $p$ gives
\begin{equation}
  \partial_p \hat{\rho}_p^{(M)}
  = \partial_p (A_M)\,\hat\rho + \partial_p (B_M) S(\hat\rho) + \partial_p(C_M) \hat\id_4.
  \label{eq:drho}
\end{equation}

Taking into account Eq.\eqref{eq:coefficients_rho_twoqubits_Mapplications}, we obtain the derivative of the final state as
\begin{equation}
  \partial_p \hat{\rho}_p^{(M)}
  =
  \begin{pmatrix}
    d_{00} & 0      & 0      & d_{03} \\
    0      & d_{11} & 0      & 0      \\
    0      & 0      & d_{22} & 0      \\
    d_{30} & 0      & 0      & d_{33}
  \end{pmatrix},
  \label{eq:drho_matrix}
\end{equation}
where
\begin{equation}
\begin{aligned}
  d_{00} &= M\xi_2(p)^{M-1}\bigl[(1-2\mu)-\xi_2(p)^M\bigr],\\
  d_{33} &= M\xi_2(p)^{M-1}\bigl[(2\mu-1)-\xi_2(p)^M\bigr],\\
  d_{11} &= d_{22} = M\xi_2(p)^{2M-1},\\
  d_{03} &= d_{30} = -4M\xi_2(p)^{2M-1}\,\sqrt{\mu (1- \mu)}.
\end{aligned}
\label{eq:d00-d33-d11-d03}
\end{equation}

The derivative $\partial_p \hat\rho_p^{(M)}$ inherits the same block structure as the final state in Eq.~\eqref{eq:rhoout}. Specifically, the subspace spanned by $\{ \ket{00}, \ket{11} \}$ forms a coupled $2\times2$ block through the off-diagonal elements $d_{03}$ and $d_{30}$, whereas the subspace spanned by $\{\ket{01},\ket{10}\}$ remains diagonal with equal entries. There is no matrix element connecting these two subspaces, implying the absence of cross-block contributions in the evaluation of the QFI through Eq.~\eqref{eq:master}.

We next determine the spectral decomposition of $\hat\rho_p^{(M)}$ Eq.~\eqref{eq:rhoout}. The subspace spanned by $\{\ket{01},\ket{10}\}$ forms a diagonal $2\times2$ block, and therefore both basis states are eigenstates of $\hat\rho_p^{(M)}$ with the doubly degenerate eigenvalue
\begin{align}
\lambda_B &= B_M+C_M
= \frac{1-\xi_2(p)^{2M}}{4}.
\label{eq:lambdaB}
\end{align}
The remaining eigenvalues arise from the $({\ket{00},\ket{11}})$-subspace, whose corresponding block is
\begin{equation}
G = 
    \begin{pmatrix}
        \mu D_M + C_M & \sqrt{\mu(1-\mu)} A_M \\
        \sqrt{\mu(1-\mu)} A_M & (1-\mu) D_M + C_M
    \end{pmatrix}.
    \label{eq:new_reduced_matrix_drhopM}
\end{equation}
Diagonalizing \eqref{eq:new_reduced_matrix_drhopM} yields the eigenvalues
\begin{align}
    \lambda_{\pm} &= \frac{1 + \xi_2(p)^{2M}}{4} \pm R, 
    \label{eq:pm_eigenvalues_rhopM}
\end{align}
where
\begin{equation}
R := \frac{\abs{\xi_2(p)^M}}{2}\sqrt{ 1 - 4\mu(1- \mu)\left[1- \xi_2(p)^{2M}\right]}.
\label{eq:R}
\end{equation}
To construct the corresponding eigenvectors, we introduce a rotation angle $\theta$ that diagonalizes the matrix $G$. The angle $\theta$ is defined in terms of the matrix elements of Eq.~\eqref{eq:new_reduced_matrix_drhopM} as
\begin{equation}
    \tan (2 \theta) = \frac{2 G_{01}}{G_{00} - G_{11}} = \frac{\sqrt{\mu (1- \mu)}}{\mu - 1/2} \xi_2(p)^{M}, 
    \label{eq:rotation_angle_def}
\end{equation}
where we expressed the coefficients $A_M, B_M C_M$ (and thus, even $D_M$) as in Eq.~\eqref{eq:coefficients_rho_twoqubits_Mapplications}. For convenience, let us introduce the shorthand notation $s_\alpha := \sin(\alpha)$ and $c_\alpha := \cos(\alpha)$.
This allows us to express the two remaining eigenvectors of $\hat{\rho}_p^{(M)}$ as
\begin{equation}
\begin{aligned}
    \ket{+_\theta} &= c_\theta \ket{00} + s_\theta \ket{11}, \\
    \ket{-_\theta} &= - s_\theta \ket{00} + c_\theta \ket{11}.
\end{aligned}
\end{equation}
Expressing $\partial_p \hat{\rho}_p^{(M)}$ in the eigenbasis $\mathcal{B} =\{\ket{+_\theta}, \ket{-_\theta},|01\rangle,|10\rangle\}$ of $\hat\rho_p^{(M)}$, we find
\begin{equation}
\left[\partial_p\hat{\rho}_p^{(M)}\right]_\mathcal{B}
=
\begin{pmatrix}
d_{++} & d_{+-} & 0 & 0\\
d_{+-} & d_{--} & 0 & 0\\
0 & 0 & d_{11} & 0\\
0 & 0 & 0 & d_{11}
\end{pmatrix},
\label{eq:drhopM_rhopM_eigenbasis}
\end{equation}
where
\begin{equation}
\begin{aligned}
  d_{++} &= \bra{+_\theta} \partial_p \hat{\rho}_p^{(M)} \ket{+_\theta} = c_\theta^2d_{00}+s_\theta^2d_{33}+s_{2\theta} d_{03},\\
  d_{--} &= \bra{-_\theta} \partial_p \hat{\rho}_p^{(M)} \ket{-_\theta} =  s_\theta^2 d_{00} + c_\theta^2d_{33}- s_{2\theta} d_{03},\\
  d_{+-} &= \bra{+_\theta} \partial_p \hat{\rho}_p^{(M)} \ket{-_\theta} =  c_{2\theta} d_{03}-s_\theta c_\theta (d_{00}-d_{33}).
\end{aligned}
\label{eq:d-elements}
\end{equation}
Here, $d_{++}$ and $d_{--}$ denotes the diagonal matrix elements associated with the eigenstates $\ket{+_\theta}$ and $\ket{-_\theta}$, respectively, while $d_{+-}$ represents the corresponding off-diagonal element. The quantity $d_{11}$ corresponds to the degenerate subspace spanned by $\{\ket{01},\ket{10}\}$. Substituting these matrix elements together with the eigenvalues $\lambda_{\pm}$ and $\lambda_B$ into the spectral representation of the QFI given in Eq.~\eqref{eq:master} yields the closed form expression
\begin{equation}
   J\left( \hat{\rho}_p^{(M)} \right)
    = \frac{d_{++}^2}{\lambda_+}
    + \frac{d_{--}^2}{\lambda_-}
    + \frac{4\,d_{+-}^2}{\lambda_++\lambda_-}
    + \frac{2\,d_{11}^2}{\lambda_B}.
  \label{eq:main}
\end{equation}
The structure of Eq.~\eqref{eq:main} admits a transparent physical interpretation. The first two terms originate from variations in the populations associated with the eigenstates $\ket{+_\theta}$ and $\ket{-_\theta}$, respectively. The third term arises from the off-diagonal matrix element $d_{+-}$ and therefore captures the contribution of quantum coherence within the $\{\ket{+},\ket{-}\}$ subspace. Finally, the fourth term is associated with the doubly degenerate subspace spanned by $\{\ket{01},\ket{10}\}$, which is generated by the action of the depolarizing channel.

Once all the expressions have been substituted into the terms of Eq.~\eqref{eq:main}, some interesting behaviors emerge.
The coherence contribution is non-vanishing only when $\sqrt{\mu(1-\mu)}\neq 0$, corresponding to a coherent superposition in the initial state. 
For increasing $M$ or increasing depolarization strength $p$, this term decays as $\sim \xi_2(p)^{4M}$, demonstrating the progressive suppression of coherence-enhanced sensitivity under repeated applications of the depolarizing channel. In contrast, the contribution from the degenerate sector scales as $\sim M^2$ in the vicinity of $p=0$, reflecting the increasing distinguishability of neighboring channel outputs as the number of channel applications increases.

\section{Slope of the RQFI for the initial state of Eq.~\eqref{eq: initial state}}
\label{sec:slope_RQFI}

In this section, we compute the derivative of the RQFI achieved by a two-qubit probe subjected to $M$ depolarizing channel applications. We assume the initial state of the probe to be the one in Eq.~\eqref{eq: initial state}. The goal is to compute the slope of the RQFI and study its behavior.  We already derived the corresponding QFI in Eq.~\eqref{eq:main}. 
Hence, recalling the definition of the RQFI in Eq.~\eqref{eq:RQFI}, we have
\begin{equation}\label{eq:product-rule}
    \partial_p \mathcal{J} \left( \hat{\rho}_p^{(M)} (\mu) \right)
    \;=\; 
    J\left( \hat{\rho}_p^{(M)} (\mu) \right)
    \;+\; p \, \partial_p J\left( \hat{\rho}_p^{(M)} (\mu) \right). 
\end{equation}

Noticing that each of the four terms of Eq.~\eqref{eq:main} has the form $T_k(p) = f_k^{2}(p)/g_k(p)$, $k \in \{\mathrm{I}, \mathrm{II}, \mathrm{III}, \mathrm{IV}\}$; their derivatives with respect to $p$ are:
\begin{equation}\label{eq:quotient}
  \partial_p T_k
  = \frac{2f_k\,\partial_p f_k}{g_k}
  - \frac{f_k^{2}\,\partial_p g_k}{g_k^{2}}.
\end{equation}
As a result, 
\begin{align}
\partial_p \mathcal{J} \left( \hat{\rho}_p^{(M)} (\mu) \right) =&
    \frac{d_{++}^2}{\lambda_+}
    + \frac{d_{--}^2}{\lambda_-}
    + \frac{4\,d_{+-}^2}{\lambda_++\lambda_-}
    + \frac{2\,d_{11}^2}{\lambda_B} \notag \\
    & + p \sum_{k=\rm I}^{\rm IV} \partial_p T_k.
    \label{eq:slope_explicit}
\end{align}
In the following, we are going to explicitly compute the expression of each term comprising \eqref{eq:slope_explicit}.

Let us start from the evaluation of the derivatives of the eigenvalues of the final density matrix $\hat{\rho}_p^{(M)}$.
The derivative of the inner-block eigenvalue in Eq.~\eqref{eq:lambdaB} reads as
\begin{equation}\label{eq:dlambdaB}
  \partial_p\lambda_B
  = M\,\xi_2(p)^{2M-1},
\end{equation}
while the derivative of the eigenvalues in Eq.~\eqref{eq:pm_eigenvalues_rhopM} is:
\begin{equation}
  \partial_p\lambda_{\pm}
  = -M\xi_2(p)^{2M-1} \pm\partial_p R, 
   \label{eq:dlambda} 
\end{equation}
where
\begin{equation}\label{eq:dR}
\partial_p R = -M \xi_2(p)^{M-1} \frac{1 - 4 \mu (1- \mu) [1 - 2 \xi_2(p)^{2M}]}{\sqrt{1 - 4 \mu (1- \mu) [1 -  \xi_2(p)^{2M}]}},
\end{equation}
where we impose $\xi_2(p)\geq0$ by restricting the range of $p \in[0, 1/2]$.

We now evaluate the derivatives of the matrix entries of $\partial_p \hat{\rho}_p^{(M)}$ of Eq.~\eqref{eq:d-elements} in the $\mathcal{B}$ basis. In particular,
\begin{align}
    \partial_p d_{++} = & c_\theta^2 \partial_p d_{00} + s_\theta^2 \partial_p d_{33} + s_{2 \theta} d_*   + 2 c_{2 \theta}  d_{03}\partial_p \theta, \notag
    \\
    \partial_p d_{--} = & c_\theta^2 \partial_p d_{33} + s_\theta^2 \partial_p d_{00} - s_{2 \theta} d_*   - 2 c_{2 \theta}  d_{03}\partial_p \theta, 
 \notag
    \\
    \partial_p d_{+-} = & c_{2 \theta} d_* + s_{\theta} c_{\theta} [\partial_p (d_{33} - d_{00}) - 4 d_{03} \partial_p \theta ],
    \label{eq:dpdpm_new}
\end{align}
where we defined $d_*:= (d_{33} - d_{00})\partial_p \theta + \partial_p d_{03}$. The derivative of the rotation angle $\theta$ can be evaluated from Eq.~\eqref{eq:rotation_angle_def} as
\begin{equation}
    \partial_p \theta = -\frac{(\mu - 1/2)\sqrt{\mu (1- \mu)}M \xi_2(p)^{M-1}}{(\mu - 1/2)^2 + \mu(1- \mu) \xi_2(p)^{2M}},
\end{equation}
while the derivatives of the matrix entries of $\partial_p \hat{\rho}_p^{(M)}$ of Eq.~\eqref{eq:d00-d33-d11-d03} in the computational basis are 
\begin{align}
  \partial_pd_{00}
  =&
  -2M(M-1)\xi_2(p)^{M-2}
  \bigl[(1-2\mu)-\xi_2(p)^{M}\bigr] \nonumber\\
  & +2M^{2}\xi_2(p)^{2M-2}, \nonumber
  \\
  \partial_pd_{33} \nonumber
  =&
  -2M(M-1)\xi_2(p)^{M-2}
  \bigl[(2\mu-1)-\xi_2(p)^{M}\bigr] \notag\\
  &
  +2M^{2}\xi_2(p)^{2M-2}, \nonumber
  \\
  \partial_pd_{03}
  = &
  8M(2M-1)\sqrt{\mu(1- \mu)}\,\xi_2(p)^{2M-2}, \nonumber
  \\
  \partial_pd_{11}
  =&
  -2M(2M-1)\xi_2(p)^{2M-2}.
  \label{eq:dd11}
\end{align}

We conclude by reporting the formal expressions of the derivatives of terms $T_{\rm I}, T_{\rm II}, T_{\rm III}, T_{\rm IV}$ with respect to $p$, which are obtained from direct application of~\eqref{eq:quotient}: 
\begin{equation}
    \begin{aligned}
        \partial_pT_{\mathrm{I}}
  =& \frac{2\,d_{++}\;\partial_pd_{++}}{\lambda_{+}}
  - \frac{d_{++}^{2}\;\partial_p\lambda_{+}}{\lambda_{+}^{2}}, \\
  \partial_pT_{\mathrm{II}}
  =& \frac{2d_{--}\;\partial_pd_{--}}{\lambda_{-}}
  - \frac{d_{--}^{2}\;\partial_p\lambda_{-}}{\lambda_{-}^{2}}, \\
  \partial_pT_{\mathrm{III}}
    =& \frac{8 d_{+-}\;\partial_pd_{+-}}{\lambda_{+}+\lambda_{-}}
   - \frac{4 d_{+-}^{2}\,\partial_p(\lambda_{+}+\lambda_{-})}{(\lambda_{+}+\lambda_{-})^{2}}, \\
   \partial_pT_{\mathrm{IV}} =& \frac{-32M^{2}\xi_2(p)^{4M-3}
            \!\left[(2M-1)-(M-1)\xi_2(p)^{2M}\right]}
          {\bigl(1-\xi_2(p)^{2M}\bigr)^{2}},
    \end{aligned}
    \label{eq:T_terms}
\end{equation}
where the quantities $d, \lambda, \partial_p d, \partial_p \lambda$ have been explicitly derived above. For the derivative of $T_{\mathrm{IV}}$, we have obtained an explicit expression in terms of $M$ and $p$. Inserting all the four derivatives of Eq.~\eqref{eq:T_terms} in Eq.~\eqref{eq:slope_explicit}, we get the full closed form of the RQFI slope with $p$, used to analyze the behavior of the RQFI in the limit $p\to 0$:
\begin{widetext}
\begin{equation}
\label{eq:final}
\begin{aligned}
  \partial_p \mathcal{J} \left( \hat{\rho}_p^{(M)} (\mu) \right)
  &= \left[
        \frac{d_{++}^{2}}{\lambda_{+}}
      + \frac{d_{--}^{2}}{\lambda_{-}}
      + \frac{4d_{+-}^{2}}{\lambda_{+}+\lambda_{-}}
      + \frac{2d_{11}^{2}}{\lambda_{B}}
     \right]
   +\; p\!\left[
      \frac{2d_{++}\;\partial_pd_{++}}{\lambda_{+}}
    - \frac{d_{++}^{2}\;\partial_p\lambda_{+}}{\lambda_{+}^{2}}
     \right]
  \\[6pt]
  &\quad +\; p\!\left[
      \frac{2d_{--}\;\partial_pd_{--}}{\lambda_{-}}
    - \frac{d_{--}^{2}\;\partial_p\lambda_{-}}{\lambda_{-}^{2}}
     \right]
 +\; p\!\left[
     \frac{8 d_{+-}\;\partial_pd_{+-}}{\lambda_{+}+\lambda_{-}}
   - \frac{4 d_{+-}^{2}\,\partial_p(\lambda_{+}+\lambda_{-})}{(\lambda_{+}+\lambda_{-})^{2}}
     \right]
  \\[6pt]
  &\quad +\; p \:
  \frac{-32M^{2}\xi_2(p)^{4M-3}}
       {\bigl(1-\xi_2(p)^{2M}\bigr)^{2}}
  \left[(2M-1)-(M-1)\xi_2(p)^{2M}\right].
\end{aligned}
\end{equation}
\label{eq:slope_complete}
\end{widetext}

\bibliography{main.bbl}

\end{document}